\documentclass[fleqn,usenatbib]{mnras}

\usepackage{newtxtext,newtxmath} 

\usepackage[T1]{fontenc} 
\usepackage{ae,aecompl} 

\usepackage{graphicx}	
\usepackage{amsmath}	
\usepackage{amssymb}	

\usepackage{breqn}  

\usepackage[normalem]{ulem}	
\usepackage{wasysym} 

\usepackage{cleveref}
\crefname{section}{\S}{\S\S}

\usepackage{etoolbox}
\makeatletter
\patchcmd\@combinedblfloats{\box\@outputbox}{\unvbox\@outputbox}{}{\errmessage{\noexpand patch failed}}
\makeatother

\newcommand{\eg}[0]{$\textnormal{e.g. }$}

\title[LBE and Ram-Pressure Stripping]{A New Method to Quantify Environment and Model Ram-Pressure Stripping in N-Body Simulations}

\author[Ayromlou et al.]{Mohammadreza Ayromlou,$^{1}$\thanks{E-mail: ayromlou@mpa-garching.mpg.de}
Dylan Nelson,$^{1}$
Robert M. Yates,$^{1}$
\newauthor
Guinevere Kauffmann,$^{1}$
Simon D. M. White$^{1}$
\\
$^{1}$Max Planck Institute for Astrophysics, Karl-Schwarzschild-Str. 1, 85741 Garching bei M{\"u}nchen, Germany}

\date{}
\pubyear{2018}

\begin{document}
\label{firstpage}
\pagerange{\pageref{firstpage}--\pageref{lastpage}}
\maketitle
\begin{abstract}
We introduce a Local Background Environment (LBE) estimator that can be measured in and around every galaxy or its dark matter subhalo in high-resolution cosmological simulations. The LBE is designed to capture the influence of environmental effects such as ram-pressure stripping on the formation and evolution of galaxies in semi-analytical models. We define the LBE directly from the particle data within an adaptive spherical shell, and devise a Gaussian mixture estimator to separate background particles from previously unidentified subhalo particles. Analyzing the LBE properties, we find that the LBE of satellite galaxies is not at rest with respect to their host halo, in contrast to typical assumptions. The orientations of the velocities of a subhalo and its LBE are well aligned in the outer infall regions of haloes, but decorrelated near halo center. Significantly, there is no abrupt change in LBE velocity or density at the halo virial radius. This suggests that stripping should also happen beyond this radius. Therefore, we use the time-evolving LBE of galaxies to develop a method to better account for ram-pressure stripping within the Munich semi-analytical model, \textsc{L-Galaxies}. Overall, our new approach results in a significant increase in gas stripping across cosmic time. Central galaxies, as well as satellites beyond the virial radius, can lose a significant fraction of their hot halo gas. As a result, the gas fractions and star formation rates of satellite galaxies are suppressed relative to the fiducial model, although the stellar masses and global stellar mass functions are largely unchanged.
\end{abstract}



\section{Introduction}
The expanding Universe is well described by the standard model of cosmology, $\rm \Lambda CDM$, on large scales. Many simulations have modelled the evolving distribution of dark matter, based on the theory of gravity with initial conditions well-specified by cosmic microwave background observations \citep{spergel2003first,komatsu2009five,ade2014planck,aghanim2018planck}. In the standard paradigm of  hierarchical structure formation theory, baryonic matter accretes into the gravitational potential wells of dark matter haloes and then cools and contracts, eventually forming stars and galaxies \citep{white1978core}. These baryons evolve under the influence of a complex set of physical processes beyond gravity alone, producing a rich phenomenology in the study of galaxy formation.

Our understanding of the formation and evolution of galaxies has improved significantly during the past few decades through detailed comparisons of theoretical models with observational data \citep[see][for reviews]{mo2010galaxy,benson2010galaxy,somerville15}. Observations provide important constraints on the distributions of morphologies, luminosities, colours, stellar masses and many other physical quantities associated with galaxies \citep[e.g.][]{stoughton2002sloan}. On the theoretical side, numerical simulations have become increasingly powerful in resolving the cosmological structure of dark matter \citep{springel2005simulations,boylan2009resolving} together with the baryonic components and their physical processes \citep{schaye10,dubois14,pillepich18b}. At the same time, the more computationally efficient approach of semi-analytical modelling has become increasing sophisticated  \citep{kauffmann1993formation,kauffmann1999clustering,somerville1999semi,cole2000hierarchical,springel2001populating,croton2006many,de2006formation}. A semi-analytical model (SAM) couples a set of equations describing key physical processes in galaxy formation to cosmological assembly histories typically extracted from `dark matter only' simulations, such as the Millennium Simulation \citep{springel2005simulations,guo2011dwarf,henriques2015galaxy}.

One fundamental aspect affecting the formation and evolution of galaxies is their environment \citep{boselli06}. It is now well known that there exists a strong dependence of galaxy properties on environment \citep{hubble1931velocity}; the morphology-density relation \citep{oemler1974systematic, dressler1980galaxy} and the enhanced quenched fractions for satellites residing within massive clusters \citep{kauffmann2004environmental,peng10,roberts19} are particularly clear examples. Tidal and ram-pressure forces can strip both star-forming gas in the disk and the hot halo gas around galaxies in dense environments \citep{gunn1972infall,binney1987galactic}. The impact of stripping on the star formation rates, colours, stellar masses, and gas contents of galaxies as a function of environment can now be quantified in large galaxy samples as well as in highly resolved data across a range of observational tracers \citep{jaffe15,poggianti17,boselli18}.

These processes have been studied with idealized hydrodynamical simulation of satellite galaxies \citep{roediger07,bekki09,tonnesen09,vijayaraghavan17}. It is also now possible to directly study gas stripping phenomena in the full cosmological context with simulations such as IllustrisTNG \citep{yun19}, in addition to the environmental impact on the gas contents of satellite galaxies more generally \citep{sales15,bahe17,wang18,stevens19}.

Early semi-analytic models assumed that hot gas was stripped out of galaxies immediately after infall into the parent halo \citep{kauffmann1993formation,de2006formation}. The \textsc{Galform} \citep{lacey2016unified} and \textsc{Shark} \citep{lagos2018shark} SAMs also include instantaneous stripping for the hot gas component of satellite galaxies, whereas \cite{font2008colours} formulated a gradual RPS process informed by hydrodynamical simulations for \textsc{Galform}. In such models there is no post-infall gas accretion onto satellites, although cold star-forming gas can continue to form stars until the galaxy eventually exhausts this reservoir \citep{larson80}.

In addition to RPS, SAMs including \textsc{Sage} \citep{croton2016semi} and \textsc{Dark Sage} \citep{stevens2016building} have also applied tidal stripping to satellite galaxies, an effect which we note is usually weaker than ram-pressure. The \textsc{Sag} semi-analytic model \citep{cora2018semi} has both tidal and ram-pressure stripping for hot gas, using the RPS method introduced in \cite{mccarthy2007ram}. Most relevant to our current effort is \cite{tecce2010ram}, who implemented a method using gas particle data from a matched hydrodynamic simulation to estimate RPS using the local density and velocity of the ICM environment through which satellite galaxies move. 

Recent versions of the \textsc{L-Galaxies} model have prescriptions for both time-evolving tidal and ram-pressure stripping of the gas within subhaloes \citep{guo2011dwarf}. An accurate model for the tidal and ram-pressure stripping of satellite galaxies requires accurate dark matter and gas profiles, such that background gas densities, as well as the velocities of satellite galaxies relative to the background gas, can be estimated. The two latter quantities are usually approximated using the properties of the satellite galaxy's host halo \citep{henriques17}. Although correct to first order, this precludes a treatment of gas stripping effects which depends on local gas inhomogeneities and structure within a host halo, as well as the stripping of central galaxies passing through cosmic environments such as filaments or sheets \citep{aragoncalvo16,kraljic18}.

The scales over which environment influences the properties of galaxies are also still a matter of discussion. Most semi-analytical models of galaxy formation and evolution, including \textsc{L-Galaxies}, adopt $R_{\rm vir}$, the virial radius, or $R_{200}$, the radius within which the matter density equals 200 times the critical density of the Universe, as the boundary for the dark matter halo and its hot gas component. In the modelling, this spatial edge acts as a sharp threshold for cutting-off environmental effects, which is clearly an oversimplified assumption. It has been noted for instance that the `splashback radius' may be a more physical boundary for a dark matter halo, as it corresponds to the radius within which at which accreted matter reaches its first orbital apocenter after turnaround. Depending on the accretion rate, the splashback radius ranges from slightly smaller than $R_{\rm vir}$ to $\approx 1.5R_{\rm vir}$ \citep{diemer2014dependence,adhikari2014splashback, more2015splashback}. Hydrodynamical simulations suggest that the shock-heated gas of a dark matter halo extends beyond its virial radius, up to $2-3R_{\rm vir}$ depending on halo mass \citep{nelson16,zinger2018quenching}.

There have been several observational \citep{hansen2009galaxy,von2010star,lu2012cfht} and theoretical \citep{balogh1999differential,bahe2012competition} studies showing that the environmental effects might extend well beyond the virial radius or similar halo boundaries. A large scale correlation between the star formation of neighboring galaxies out to distances as large as 10 Mpc has also been observed -- galactic conformity \citep{weinmann2006properties,kauffmann2013re,kauffmann2015physical,hearin2016physical,hatfield2017environmental,treyer2018group}. To capture such physical effects, a realistic semi-analytic galaxy model needs to contain prescriptions for environmental effects which are \textit{local}, and which avoid artificial boundaries.

In this work we measure local properties in the immediate vicinity of dark matter subhaloes, namely background density and bulk velocity. We use this local background environment (LBE) to devise a more realistic treatment of ram pressure stripping in the \textsc{L-Galaxies} SAM. We also investigate a variety of background properties using our LBE technique and dark matter particle data from the Millennium Simulation \citep{springel2005simulations}. 

This paper is structured as follows: In \S \ref{sec: Methodology}, we describe the simulation, the Munich Semi-Analytic Model, \textsc{L-Galaxies}, and gas stripping theory. In \S \ref{sec:LBE} we introduce the definition of local background environment (LBE) and discuss how it can be used for the calculation of ram-pressure force. Analysis of the LBE is given in \S \ref{sec: analyse_lbe}, where we consider its velocity, density, and the correlation with galaxy and subhalo properties. In \S \ref{sec: SAM}, we show the results of our new method applied to \textsc{L-Galaxies} and the resulting changes to the baryonic properties of galaxies. We conclude and summarize in \S \ref{sec: conclusions}.


\section{Methodology}
\label{sec: Methodology}

\subsection{Simulation and Subhalo Identification}
\label{subsec: MS}

In this work we use the particle and halo merger tree data of the Millennium Simulation \citep{springel2005simulations}. The Millennium Simulation (hereafter `MS') has 63 snapshots. There are about 10 billion particles ($2160^3$) in the simulation box, which has a co-moving volume of $(500 \rm Mpc/h)^3$ and a particle mass of $m_{\rm p} = 8.61 \times 10^8 \rm  \rm M_{\odot}/h$. The MS is based on the WMAP1 cosmology \citep{spergel2003first}; in this paper we rescale to an updated Planck cosmology \citep{ade2014planck} using the scaling method of \cite{angulo2010one} as updated in \cite{angulo2015cosmological}. The cosmological parameters are then: $\Omega_{\rm m}=0.315$, $\Omega_{\rm b}=0.049$, $\Omega_{\Lambda}=0.685$, $h = 0.673$ and $\sigma_{\rm 8} = 0.826$. The co-moving size of the box then becomes $(480 \rm Mpc/h)^3\approx (714 \rm Mpc)^3$ and the particle mass $9.6\times 10^8 \rm  \rm M_{\odot}/h$. Redshift zero occurs at the 58th snapshot, and snapshots 59-63 represent the future.

Dark matter haloes are identified in all snapshots using a Friends Of Friends (FOF) algorithm. Each FOF halo has one central subhalo, while the others are labeled as satellite subhaloes. All such subhaloes are identified with the \textsc{Subfind} algorithm \citep{springel2001populating}. We note that satellite subhaloes of a FOF halo can exist both within and beyond $R_{200}$. The minimum number of particles for a subhalo to be included in the catalogues is set to be 20.

\subsection{The L-Galaxies Semi-Analytical Model}
\label{sec: munich sam}

We use the Munich semi-analytical model (hereafter `Munich SAM'), and the latest version of its publicly available code, \textsc{L-Galaxies}\footnote{http://galformod.mpa-garching.mpg.de/public/LGalaxies} \citep[][hereafter `H15']{henriques2015galaxy}, to model the formation and evolution of galaxies, including environmental effects. \textsc{L-Galaxies} is a self-consistent semi-analytical model of galaxy formation. It contains a set of simplified physical recipes to describe processes such as gas cooling, star formation, feedback from supernovae and active galactic nuclei and predicts the properties of a large ensemble of galaxies by implementing these recipes on dark matter subhalo merger trees generated from a dark matter N-body simulation. 

The method starts by assigning diffuse hot gas based on the cosmic baryon fraction to each subhalo at its formation redshift and then follows cooling processes that produce cold star-forming gas from which stars are born. Energy released during the death of stars reheats the cold gas and pushes it into the hot halo atmosphere. A fraction of the heated gas can itself be removed into an ejected reservoir, which returns to the galaxy at a later snapshot.

There are various physical mechanisms which can quench star formation, including supernova feedback, AGN feedback and environmental effects such as ram-pressure and tidal stripping. Starting from \cite{guo2011dwarf} (hereafter `Guo11') \textsc{L-Galaxies} implements a model for tidal and ram-pressure stripping of satellite galaxies within $R_{200}$. As a result, satellite galaxies can retain a hot gas reservoir post-infall. However, the inclusion of RPS in low mass haloes led to an excessive satellite quenched fraction. H15 therefore artificially limited ram-pressure stripping (RPS) to satellites of massive haloes only ($>2\times 10^{14} \rm M_\odot/h$) in order to reproduce the colour distributions of satellites as a function of host mass. Taking H15 as our base model we first eliminate this halo mass threshold and then proceed to update the RPS model itself. Detailed descriptions of physical processes in the base model can be found in \cite{guo2011dwarf} and the supplementary material of \cite{henriques2015galaxy}.

In \textsc{L-Galaxies}, a galaxy's `type' is defined according to its subhalo status in the Friends Of Friends (FOF) group. The galaxy associated with the central subhalo of the FOF halo is the central or type 0 galaxy. All other galaxies residing in resolved dark matter subhaloes in the same FOF halo are called type 1 satellite galaxies. Finally, type 2 satellites (orphan galaxies) have subhaloes which have passed below the resolution limit and may be completely disrupted. The FOF algorithm links particles at fixed inter-particle separation and the resulting FOF system can have a non-spherical shape, meaning that there is no strict correspondence requiring satellite subhaloes to be within $R_{200}$.

We define $M_{\rm 200}$ and $R_{\rm 200}$ as the mass and radius of a FOF halo at an overdensity of 200 times the critical density of the universe. We take these two values as the virial radius $R_{\rm vir}$ and virial mass $M_{\rm vir}$, respectively. Furthermore, we define $M_{\rm 200,max}$ as the maximum virial mass over the history of a subhalo. The subhaloes of satellite galaxies have lost mass through tidal stripping after falling into a larger system, so $M_{\rm 200,max}$ will generally correspond to the maximum $M_{\rm 200}$ at the time they were last a central galaxy.

In the rest of this paper, we select galaxies above a stellar mass limit of $M_\star = 3\times10^9 \rm M_{\odot}/h$  unless stated otherwise. This is the mass limit above which the \textsc{L-Galaxies} stellar mass functions of the Millennium and higher resolution Millennium-II simulations are converged \citep{guo2011dwarf}. This mass limit ensures that our results are, as much as possible, robust to resolution. There are over two million galaxies above this mass limit in the simulation volume at $z=0$, of which about 1.3 million are categorized as centrals and 0.7 million as satellite galaxies (0.5 million as type 1, and 0.2 million as type 2).

\subsection{Hot gas stripping}
\label{sec: rps and tidal}

\subsubsection{Ram-pressure stripping}

Ram-Pressure Stripping (RPS) can act on both the hot and the cold gas components of a galaxy. Physical prescriptions for the RPS of hot gas are currently included in the fiducial \textsc{L-Galaxies} model. Techniques to include ram-pressure stripping of cold gas in disks have also been explored \citealt{Luo2016Resolution}), but these are not yet included in the publicly-released version of the model and we do not adopt them here. In this paper, we focus on a more accurate model for the stripping of the hot gas. We adopt a similar approach as in \cite{mccarthy2007ram,guo2011dwarf,henriques2015galaxy} to derive the ram-pressure stripping radius, the radius outside which all hot gas is assumed to be stripped. Our implementation contains a new method to resolve environmental properties as well as an updated estimation of the self-gravity of galaxies.

In general, gas can be stripped out of a subhalo if the ram-pressure force from its environment overcomes the gravitational force on its gas component. The scale on which that happens is defined by the ram-pressure stripping radius, $R_{\rm rp}$. Beyond the stripping radius, all the hot gas is assumed to be stripped. The ram-pressure can be calculated using the formula of \cite{gunn1972infall} as

\begin{equation}
\label{eq: rp_rp}
P_{\rm rp} = \rho_{\rm LBE,gas} \, v_{\rm gal,LBE}^2 \,,
\end{equation}

\noindent where $\rho_{\rm LBE,gas}$ is the gas density of the galaxy's local background environment (LBE, see \S \ref{sec:LBE_Definition}) and $v_{\rm gal,LBE}$ is the velocity of the galaxy relative to the environment it is moving through. The gravitational force per unit area between the galaxy's hot gas and its subhalo at a given radius from the center of the subhalo is

\begin{equation}
\label{eq: rp_grav}
F_{\rm g}(r) = g_{\rm max} \,\rho^{\rm proj}_{\rm hotgas} \,,
\end{equation}

\noindent where $\rho^{\rm proj}_{\rm hotgas}(r)$ is 2D projected hot gas density around the galaxy and $g_{\rm max}$ is the maximum restoring gravitational field in the direction of galaxy's velocity relative to its LBE.

Calculation of the subhalo mass within a given radius requires assumptions for the density profiles of the dark matter and gas components. We assume, as before, subhalo and hot gas density profiles to be isothermal spheres with $\rho \propto r^{-2}$. Therefore, the maximum gravitational field is

\begin{equation}
g_{\rm max} = \frac{GM_{\rm subhalo}(r)}{2r^2},
\end{equation}

\noindent where $M_{\rm subhalo}(r)$ is the  subhalo mass within the radius $r$ and is defined as

\begin{equation}
\label{eq: M_subhalo_rp}
M_{\rm subhalo}(r) = M_{\rm g}\frac{r}{R_{\rm g}},
\end{equation}

\noindent where $R_{\rm g}$ and $M_{\rm g}$ are the radius and mass within which we estimate the gravitational potential of the subhalo, respectively. For central galaxies, we take $R_{\rm g}=R_{\rm 200}$ and $M_{\rm g}=M_{\rm 200}$. For satellite galaxies, $R_{200}$ and $M_{200}$ are not appropriate values. Previous \textsc{L-Galaxies} models (e.g. \citealt{guo2011dwarf,henriques2015galaxy}) used the mass and radius at the time when the satellite galaxy was last a central, i.e. $M_{\rm 200,infall}$ and $R_{\rm 200,infall}$.  Here we take $R_{\rm g}$ to be the half-mass radius of the subhalo, $R_{\rm halfmass}$, and $M_{\rm g}$ to be the total mass within half-mass radius at its current snapshot. The mass contained within this radius should constitute a more faithful representation of the actual gravitational potential of the subhalo at late times, once significant tidal stripping has occurred.

Adopting an isothermal profile with $\rho \propto r^{-2}$ for the galaxy's hot gas halo, the 2D projected hot gas density can be estimated as

\begin{equation}
\rho^{\rm proj}_{\rm hotgas}(r) = \frac{M_{\rm hotgas}}{2 \pi R_{\rm hotgas}r},
\end{equation}

\noindent where $M_{\rm hotgas}$ and $R_{\rm hotgas}$ are the hot gas mass and radius of the subhalo which undergoes RPS, and the coefficient $1/2\pi$ is chosen so that the integral of the projected density equals $M_{\rm hotgas}$ within projected radius $R_{\rm hotgas}$.

Stripping occurs when ram-pressure overcomes gravity. The radius within which this happens is called the ram-pressure stripping radius, $R_{\rm rp}$, and is given by

\begin{equation}
R_{\rm rp} = \left(\frac{GM_{\rm g}M_{\rm hotgas}}{4\pi R_{\rm g} R_{\rm hotgas} \rho_{\rm LBE,hotgas} v_{\rm gal,LBE}^2} \right)^{1/2}.
\end{equation}

\noindent Given the two main properties of a galaxy's LBE (i.e. the density and the velocity of the environment through which the galaxy is moving), the above estimate for the stripping radius can be applied to all of the galaxies in the simulation.

For reference, we note that \citet{guo2011dwarf} and \citet{henriques2015galaxy} considered RPS only on satellite galaxies within $R_{200}$. For each satellite galaxy, they adopted a local gas density assuming an isothermal gas density profile ($\rho\propto r^{-2}$), and took the velocity of galaxy relative to its LBE, $v_{\rm gal,LBE}$, as the virial velocity of the host FOF halo. These are reasonable average estimates, but they are not local measurements, and do not extend to satellites beyond the virial radius or to central galaxies.

\subsubsection{Tidal Stripping}

In addition to RPS, tidal stripping also removes gas from satellite galaxies. Typically, SAMs assume that the fraction of stripped hot gas equals the fraction of dark matter lost by the subhalo \citep[e.g.][]{guo2011dwarf,henriques2015galaxy}. Therefore,

\begin{equation}
\frac{M_{\rm hot}(R_{\rm tidal})}{M_{\rm hot,infall}} = \frac{M_{\rm DM}}{M_{\rm DM,infall}}.
\end{equation}

\noindent For our assumed isothermal density profile, this gives the following expression for the tidal stripping radius

\begin{equation}
R_{\rm tidal} = \frac{M_{\rm DM}}{M_{\rm DM,infall}}R_{\rm hot,infall} \,,
\end{equation}

\noindent where $M_{\rm DM}$ is the current mass of the satellite, $M_{\rm DM,infall}$ is its virial mass at infall and $R_{\rm hot,infall}$ is its hot gas radius at infall which is assumed to be $R_{\rm 200,infall}$, since for central galaxies the hot gas radius is set to $R_{200}$. We only apply tidal stripping to satellite galaxies.

\subsubsection{Stripping implementation}

For central galaxies, the stripping radius is equal to the ram-pressure stripping radius. For satellite galaxies, however, we take the stripping radius $R_{\rm strip}$ to be the minimum of the tidal and ram-pressure stripping radii. In general, if $R_{\rm strip} < R_{\rm hotgas}$, the gas beyond $R_{\rm strip}$ will be lost. After stripping, the density profile of the remaining hot gas is assumed to remain a truncated isothermal.

Each galaxy in \textsc{L-Galaxies} has an ejected reservoir of gas in addition to its cold and hot gas reservoirs. The mass in the ejected reservoir can either return to the hot gas reservoir or be stripped because of ram-pressure and tidal forces. In this paper, we take the fraction of mass stripped from the ejected reservoir to be equal to the mass fraction stripped from the hot reservoir. The hot gas stripped from satellites is added to the central hot gas reservoir, and the stripped ejected reservoir from satellites is added to the central ejected reservoir. This treatment is the same as in \citet{guo2011dwarf} and \citet{henriques2015galaxy}. For central galaxies, gas stripped (from either the hot or ejected reservoirs) is placed in a stripped reservoir, and mass in the stripped reservoir never returns to the galaxy.


\section{Local Background Environment}
\label{sec:LBE}
\subsection{Definition}\label{sec:LBE_Definition}

We define the Local Background Environment (LBE) of each galaxy within a specifically defined spherical shell surrounding its subhalo. The radii of the shell are chosen to exclude the galaxy and its subhalo (see \S \ref{sec:LBE_shell_definition}). The density of the LBE, $\rho_{\rm LBE}$, is the number density of simulation particles within this spherical shell multiplied by the particle mass. The nett velocity of the LBE, $\vec{v}_{\rm LBE}$ is the mean velocity of these shell particles.

Removing the galaxy and its subhalo from the LBE estimate is critical for our study. Environmental effects such as RPS are not be caused by the galaxy itself; they occur due to a galaxy's movement through its environment, so we need to separate the galaxy/subhalo system from its environment in a clean way. In a simulation, however, the background shell can contain two species of particles: true LBE particles, and contaminating particles that are associated to the galaxy's own subhalo. The latter are removed, as described below, before deriving the LBE properties.

\subsubsection{Defining the background shell size}
\label{sec:LBE_shell_definition}

For galaxies with an identified subhalo, i.e. centrals and type 1 satellites, we choose the inner and outer radii of the shell to be $R_{\rm in} = \alpha_{\rm in} R_{\rm subhalo}$ and $R_{\rm out} = \alpha_{\rm out} R_{\rm subhalo}$, where $R_{\rm subhalo}$ is the subhalo size and is defined as the distance between the most bound and most distant particles of the subhalo, as identified by \textsc{Subfind}. $\alpha_{\rm in}$ and $\alpha_{\rm out}$ are coefficients related to the scale of interest for our LBE estimates.

To avoid contaminating the LBE with dark matter from its own subhalo, $\alpha_{\rm in}$ must be greater than 1. In practice, we choose $\alpha_{\rm in} = 1.25$ to further reduce contamination, making the measurement slightly less compact as a trade-off. In \S \ref{subsec: Gaussian_mixture}, we discuss the remaining contamination in more detail, and introduce a method to remove it effectively.

The outer radius of the background shell is chosen to be the maximum of $2R_{\rm subhalo}$ and the radius which encompasses a minimum number of particles, $n_{\rm min}$ in the shell. We choose $n_{\rm min}=30$, which gives a statistical error of less than $n_{\rm min}^{-1/2} \approx 20\%$. For more than half of the subhaloes in MS, $\alpha_{\rm out} = 2$ satisfies the $n_{\rm min} = 30$ limit, and less than 1\% of subhaloes reach $\alpha_{\rm out} > 4$; these reside in very low density regions. The typical number of shell particles around satellite galaxies in groups and clusters ranges between a few hundred and a few thousand. Increasing $n_{\rm min}$ much beyond our fiducial choice would force the outer shell radius out to an unacceptably large distance in low density regions for little gain in the accuracy of the local density estimate.

Type 2 satellites (orphan galaxies) are not subject to RPS for hot gas since they have already lost their subhalo along with its hot gas component. However, to make our analysis comprehensive, we measure their LBE properties as well, choosing $R_{\rm in} = 0$. Because type 2 satellites are usually found near the center of FOF haloes (see Fig. \ref{fig: types_dis_pdf}), an accurate density estimate requires a relatively small outer radius. We therefore set $R_{\rm out}$ to be the maximum of $0.04R_{\rm 200,host}$ and the radius which encompasses $n_{\rm min} = 30$ particles. Increasing/decreasing the outer radius by a factor of 2 does not change the properties of type 2 satellite LBE significantly. Similar to central and type 1 satellite galaxies, there will also be contaminating particles from the galaxy's subhalo when it was last resolved, and we remove this component as described in \S \ref{subsec: Gaussian_mixture}.

\subsection{Removing LBE contamination using a Gaussian Mixture Method (GMM)}
\label{subsec: Gaussian_mixture}

\begin{figure*}
\centering
  \includegraphics[width=1\textwidth]{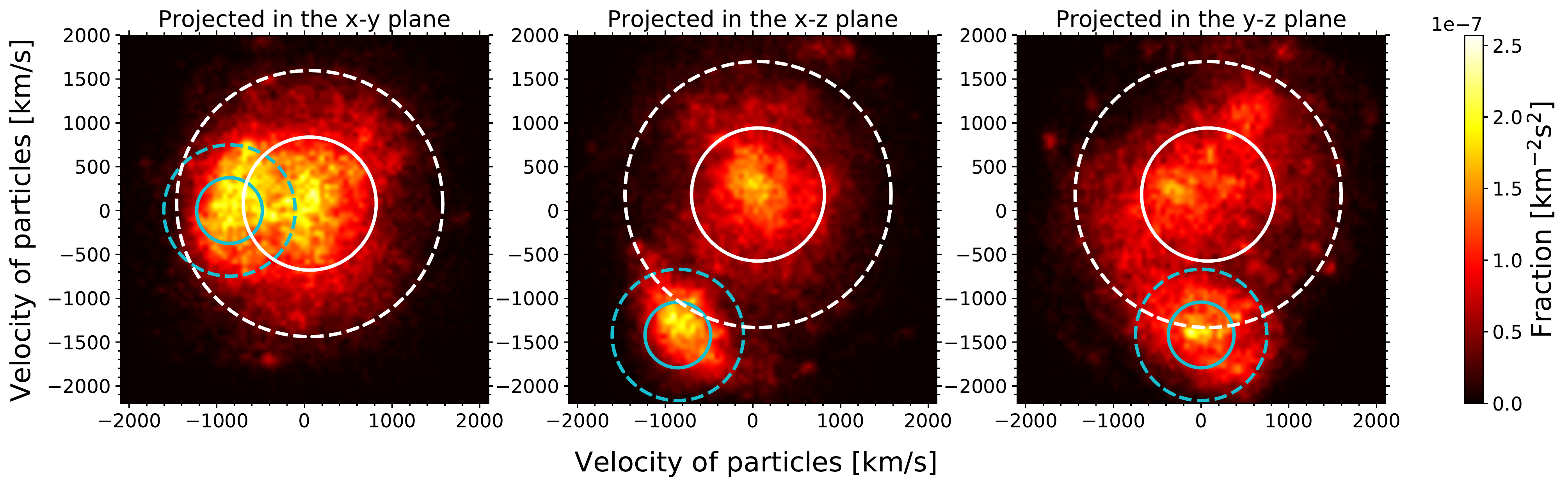}
  \includegraphics[width=1\textwidth]{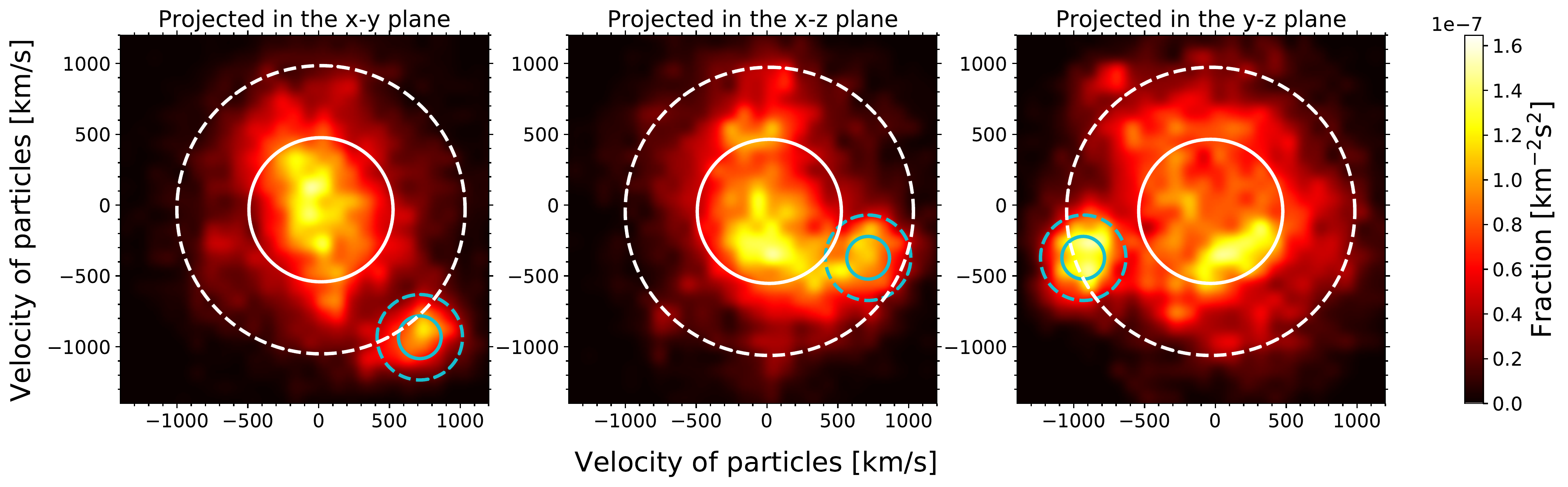}
  \caption{Distribution of the velocity of particles in the background shell of two satellite galaxies (each row shows one galaxy), containing 60,000 and 8,000 particles, respectively. The velocities are reported in the rest frame of the satellite galaxy's host halo. The colours show the fraction of particles in each velocity pixel. The cyan circles are centered at the velocity of the galaxy, and the white circles are centered at the derived mean velocity of the LBE after
  decontamination. The radius of solid circles is equal to velocity dispersion of the two modelled Gaussians, while dashed circles show twice that value. The fraction of contaminant particles in the top panel is 0.22 and 0.10 in the bottom panel. The three panels in each row correspond to three different projections of the 3D velocity distribution in the x-y, x-z and y-z planes. The magnitude and direction of the LBE velocity vectors are strongly affected by our decontamination procedure.}
\label{Fig: velpar_pdf_2d}
\end{figure*}

Removing all the subhalo particles which are identified by the \textsc{Subfind} algorithm does not result in a background shell being completely emptied of subhalo-associated particles. These near-members may be on the edge of gravitational boundedness, and typically move with a velocity that is close to the velocity of the subhalo itself, but which is significantly different than true background particles in the shell. Their classification as members of the subhalo is largely a matter of definition, i.e. a detail of the \textsc{Subfind} algorithm, but for the purposes of classifying the true LBE, we aim to identify them as part of the subhalo itself.

We adopt a model in which the background shell of each subhalo is formed by two different species of particles. The first species consist of particles with approximately the same mean velocity and velocity dispersion as the subhalo. The second species, the true LBE particles, have unknown mean velocity and velocity dispersion. 


We wish to derive the fraction of each particle species and hence the density of each species in the shell, as well as the mean velocity (3 dimensional) and 1D velocity dispersion of the LBE particles. We consider the distribution of the velocity of each species to be a Gaussian, with known mean and dispersion for the subhalo particles, and unknown mean and dispersion for LBE particles. We write the distribution of the velocities of shell particles as a Gaussian mixture

\begin{multline}
\label{eq: Gaussian_mixture}
P(v) = f_{\rm sub} \,(2\pi\sigma_{\rm subhalo})^{-3/2} \,\exp\left( -\frac{|\vec{v} - \vec{v}_{\rm subhalo}|^2}{2\sigma_{\rm subhalo}^2} \right) \\
+ (1-f_{\rm sub}) \,(2\pi\sigma_{\rm LBE})^{-3/2} \,\exp\left( -\frac{|\vec{v} - \vec{v}_{\rm LBE}|^2}{2\sigma_{\rm LBE}^2} \right). \hspace{3.8em}
\end{multline}

Here, $\vec{v}_{\rm subhalo}$ and $\vec{v}_{\rm LBE}$ are the mean velocities of the subhalo and its LBE, while $\sigma_{\rm subhalo}$ and $\sigma_{\rm LBE}$ are their corresponding 1D velocity dispersion.  $f_{\rm sub}$ is the fraction of particles in the background shell that belong to the subhalo. Eq. \ref{eq: Gaussian_mixture} contains 5 unknown variables, $f_{\rm sub},\vec{v}_{\rm LBE}$ (3 components) and $\sigma_{\rm LBE}$. To proceed, we write down the first and second moments of the velocity of shell particles

\begin{equation}
\label{eq: first_moment}
<\vec{v}_{\rm shell}> = f_{\rm sub}<\vec{v}_{\rm subhalo}> + (1-f_{\rm sub})<\vec{v}_{\rm LBE}> \,,
\end{equation}

\begin{equation}
<|\vec{v}_{\rm shell}|^2> = f_{\rm sub}<|\vec{v}_{\rm subhalo}|^2> + (1-f_{\rm sub})<|\vec{v}_{\rm LBE}|^2>.
\end{equation}

\noindent In these four equations, $<\vec{v}_{\rm shell}>$ and $<|\vec{v}_{\rm shell}|^2>$ can be computed directly from the shell particles, while the values of $<\vec{v}_{\rm subhalo}>$ and $<|\vec{v}_{\rm subhalo}|^2>$ are known. This leaves us with 4 equations and 5 unknowns. To solve for $f_{\rm sub}$ we use a maximum log-likelihood method for the velocity distribution of shell particles according to Eq. \ref{eq: Gaussian_mixture}. The log-likelihood can be written as
 
\begin{equation}
L = \displaystyle\prod_{\rm i=1}^{\rm N} P(v_{\rm i}) \quad \Rightarrow \quad \ln L = \displaystyle\sum_{\rm i=1}^{\rm N} \ln P(v_{\rm i}),
\end{equation}

\noindent where the sum is over all of the shell particles ($N$ particles in total). Enforcing the constraint that $0\leq f_{\rm sub} \leq 1$, we numerically maximize the likelihood via a grid search and derive the fraction of subhalo particles for each shell. 


Given $f_{\rm sub}$, we can calculate the velocity, velocity dispersion, and density of the LBE. We define the shell density as the average density of the shell

\begin{equation}
\rho_{\rm shell} = \frac{1}{V_{\rm shell}}\displaystyle\sum_{\rm i=1}^{N} m_{\rm i} \quad , \quad
V_{\rm shell}=\frac{4}{3}\pi(R_{\rm out}^3 - R_{\rm in}^3),
\end{equation}

\noindent where $m_{\rm i}$ is the mass of each particle inside the shell and $V_{\rm shell}$ is the shell volume. Hence, the LBE density is

\begin{equation}
\rho_{\rm LBE} = (1-f_{\rm sub})\rho_{\rm shell}.
\end{equation}

In addition, the mean LBE velocity is driven from Eq. \ref{eq: first_moment}, and is
\begin{equation}
\vec{v}_{\rm LBE} = \frac{<\vec{v}_{\rm shell}> - f_{\rm sub}\vec{v}_{\rm subhalo}}{1-f_{\rm sub}}
\end{equation}

\noindent Finally, the velocity of a galaxy relative to its LBE becomes

\begin{equation}
\vec{v}_{\rm gal,LBE} = \vec{v}_{\rm gal} - \vec{v}_{\rm LBE},
\end{equation}

\noindent where $\vec{v}_{\rm gal}$ is the velocity of the galaxy, as determined by the average over its constituent subhalo particles. For type 2 satellites, which have no subhalo, $\vec{v}_{\rm gal}$ is the current velocity of the galaxy, which is identified with the most bound particle of its subhalo at the last time this was identified.

Fig. \ref{Fig: velpar_pdf_2d} visualizes the 2D projected velocity distribution of particles in the background shell of two different type 1 satellite galaxies. Each row shows one galaxy, and the three panels in each row present projections in the x-y, x-z and y-z planes. The component mixture can clearly be seen for both galaxies. The circles are centered at the Gaussian mixture method's predictions for the velocities of the galaxy (cyan circles) and its LBE (white circles). The solid circles show the velocity dispersion of each particle species, which broadly encloses the most populated velocity bins. The two cases presented in Fig. \ref{Fig: velpar_pdf_2d} have contamination values of $f_{\rm sub}= 0.22$ and $f_{\rm sub} = 0.10$.

To illustrate the influence of decontaminating the background shell we compare the velocity of the shell for the galaxy in the top panel of Fig. \ref{Fig: velpar_pdf_2d} before and after decontamination. In the rest frame of its host FOF halo, the mean velocities in each of the three Cartesian directions of this satellite's background shell before decontamination are $\vec{v}_{\rm shell} \simeq (-142,62,-170)$ km/s. After decontamination, the mean velocities of the LBE are $\vec{v}_{\rm LBE} \simeq (59,79,183)$ km/s. This strong difference between shell velocity and LBE velocity, both in magnitude and orientation, shows the importance of decontaminating the background shell.

\begin{figure}
\centering
 \includegraphics[width=1\columnwidth]{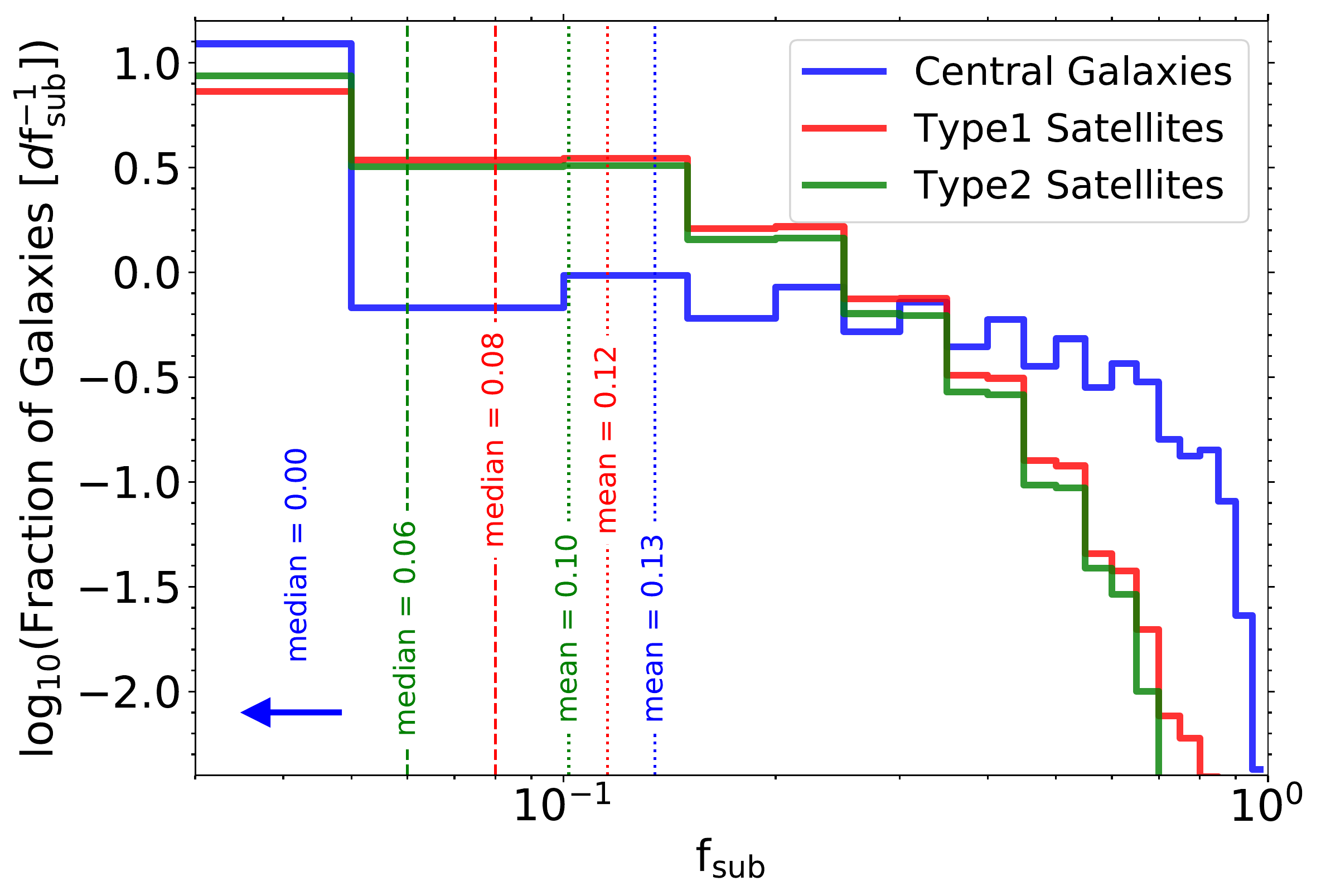}
 \caption{Distribution of the fraction of contaminant particles (subhalo particles), $f_{\rm sub}$, in the background shell of galaxies. The blue lines illustrate central galaxies, while red and green lines correspond to type 1 and type 2 (orphan) satellite galaxies. Solid lines show the fraction of contaminants divided by the size of each bin. Dashed and dotted lines denote median and mean values of $f_{\rm sub}$, which are also labeled with their value in the figure.}
\label{Fig: hist_fsub}
\end{figure}

To provide a sense of the level of contamination in our background shells, the average and median values derived for $f_{\rm sub}$ along with its distribution are shown in Fig. \ref{Fig: hist_fsub}. The y-axis illustrates the fraction of galaxies divided by the size of each bin and the x-axis denotes the value of $f_{\rm sub}$. The substantial difference between mean and median values for central galaxies reflects the large tail to high contamination fraction. Half of all central galaxies have $f_{\rm sub} = 0$. In contrast, in type 1 satellites, the median and mean values for $f_{\rm sub}$ are significantly closer to each other and are around 0.1. For type 2 satellite galaxies, however, the contamination is not as strong as type 1 satellites.

\subsection{The Density and Velocity of LBE Hot Gas}

The LBE represents the total matter background density around galaxies. Therefore, it must be translated into a hot gas density for the purpose of computing the ram-pressure stripping. For central galaxies, we multiply the total density within the LBE shell by the cosmic baryon fraction $\Omega_{\rm b}$. For satellite galaxies, the hot gas density is taken to be the total LBE density multiplied by the hot gas fraction of the associated main subhalo of the their FOF halo:

\begin{equation}
\begin{aligned}
  \rho_{\rm LBE,hotgas} &= \Omega_{\rm b} \,\rho_{\rm LBE} \qquad &\textrm{(Central Galaxies)} \\
  \rho_{\rm LBE,hotgas} &= f_{\rm hotgas} \,\rho_{\rm LBE} \qquad &\textrm{(Satellite Galaxies)},
\end{aligned}
\end{equation}

\noindent where $f_{\rm hotgas}$ is the hot gas fraction of the main subhalo of the halo hosting the satellite galaxy. In both cases we therefore assume that gas traces dark matter linearly, which is what most of the semi-analytic models (including \textsc{L-Galaxies}) assume in order to model the diffuse baryon components of haloes and subhaloes.

\begin{figure*}
  \includegraphics[width=1\textwidth]{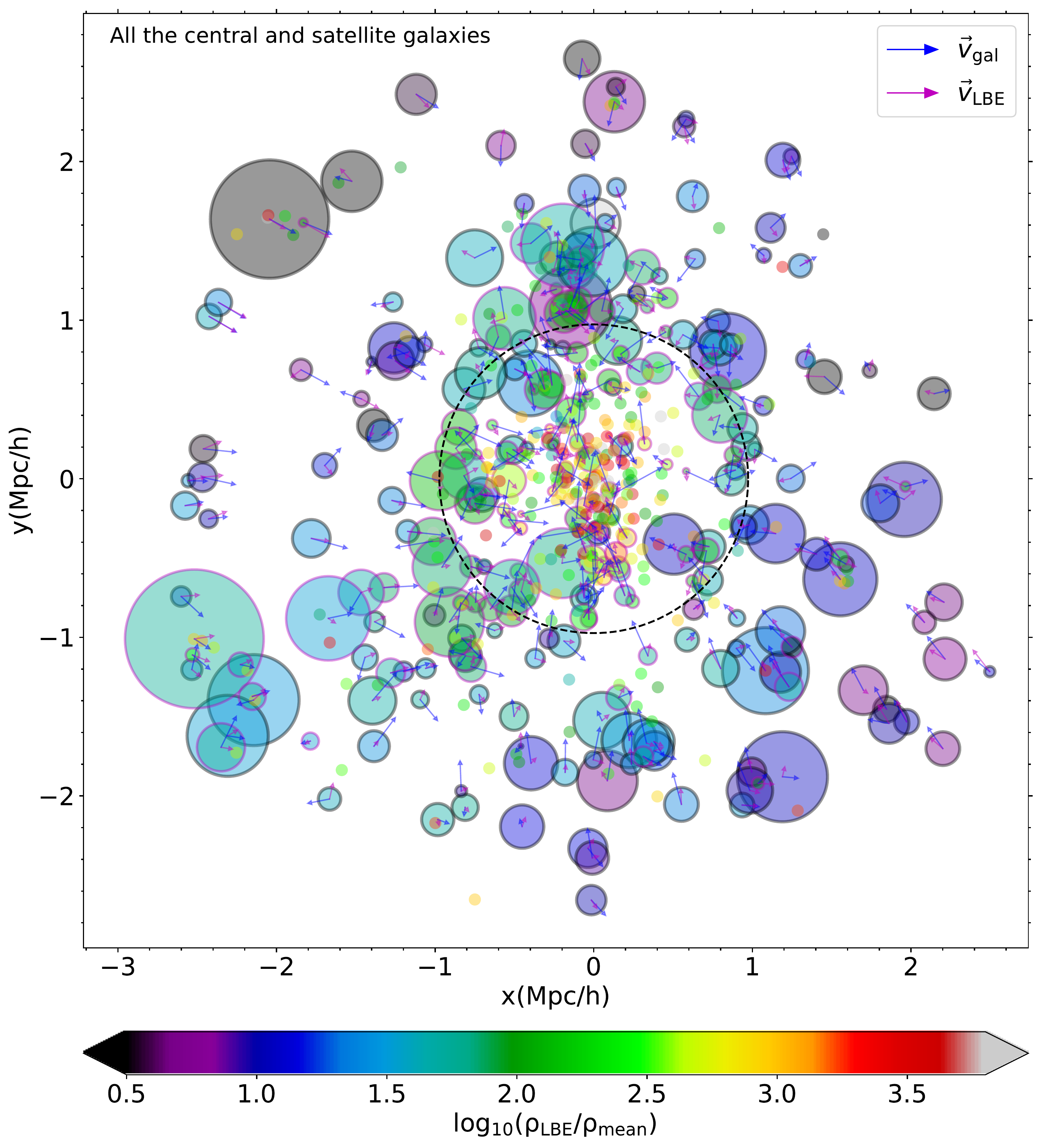}
  \caption{Environmental properties of the galaxies within $3\times R_{\rm 200}$ of a massive halo with $M_{\rm 200}\approx 10^{\rm 14} \rm M_{\odot}/h$ at redshift zero. Each circle shows a galaxy, and there are in total 582 galaxies visible, of which 169 are centrals, 120 are type 1 satellites and 293 are type 2 satellites. We note that the stellar mass limit, $M_\star = 3 \times 10^9 \rm M_{\odot}/h$, is not applied in this figure. The face-colour of each circle corresponds to the local background density (in the units of mean density of the universe) and its edge colour specifies the galaxy's type: black for centrals (type 0), red for type 1 satellite, and no edge colour for type 2 satellites. The size of circles is equal to the subhalo size for type 0 and type 1 galaxies, while type 2 satellite galaxies are simply shown as dots. The dashed black circle around the center corresponds to $R_{200}$ for the main halo. The arrows illustrate the velocities for type 0 and type 1 galaxies, where the length of each arrow is proportional to the magnitude of the velocity. The length of the arrows shown in the key is $0.1 {\rm Mpc/h}\equiv 600 {\rm km/s}$. Blue arrows show galaxy velocities in the rest frame of the main halo, and the magenta arrows show the local background velocities in the rest frame of the main halo. For type 2 galaxies, the velocities are plotted separately in Fig. \ref{Fig: BGEnv_Schematic21}.}
\label{Fig: BGEnv_Schematic}
\end{figure*}

\begin{figure*}
  \includegraphics[width=1\columnwidth]{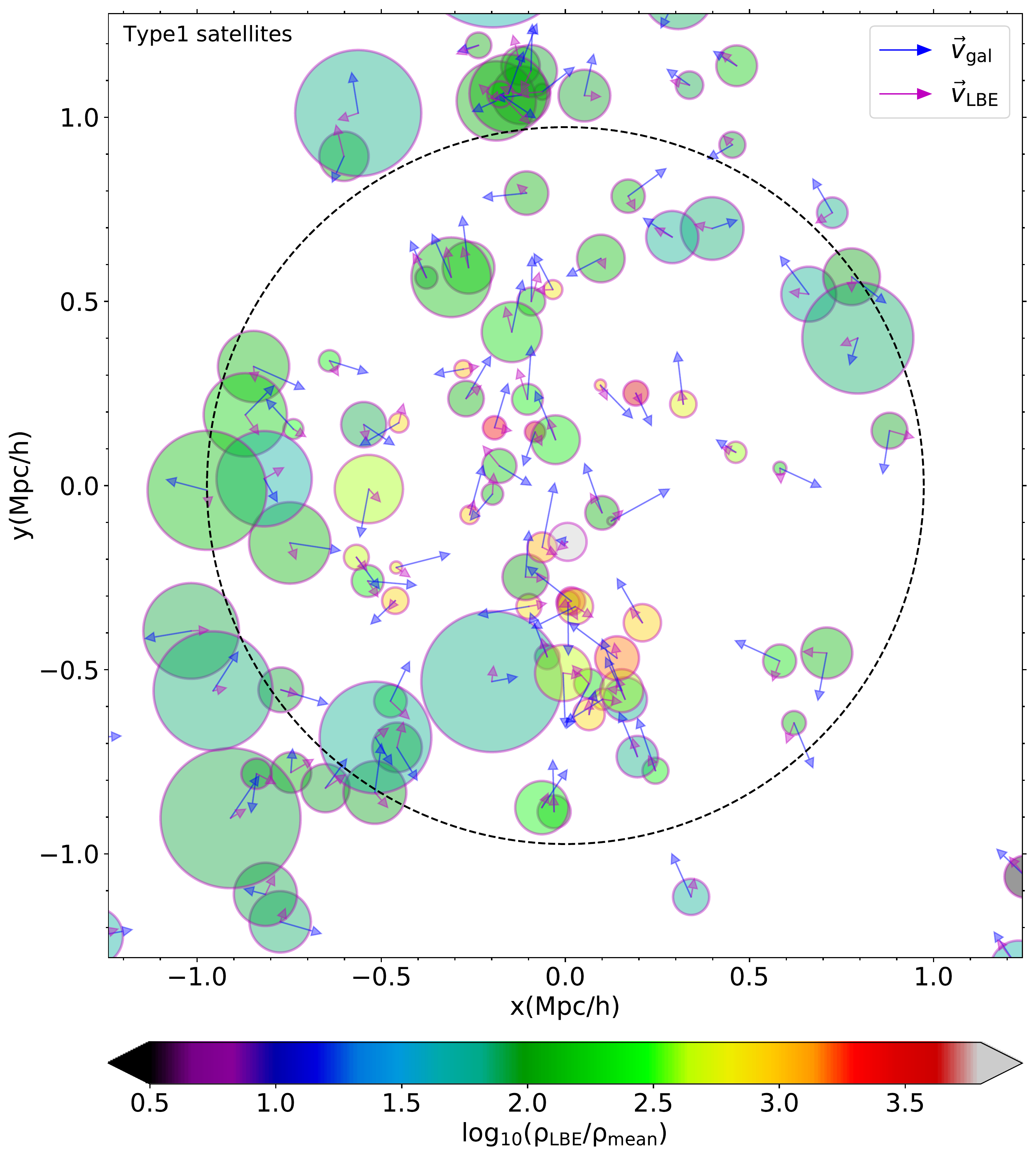}
  \includegraphics[width=1\columnwidth]{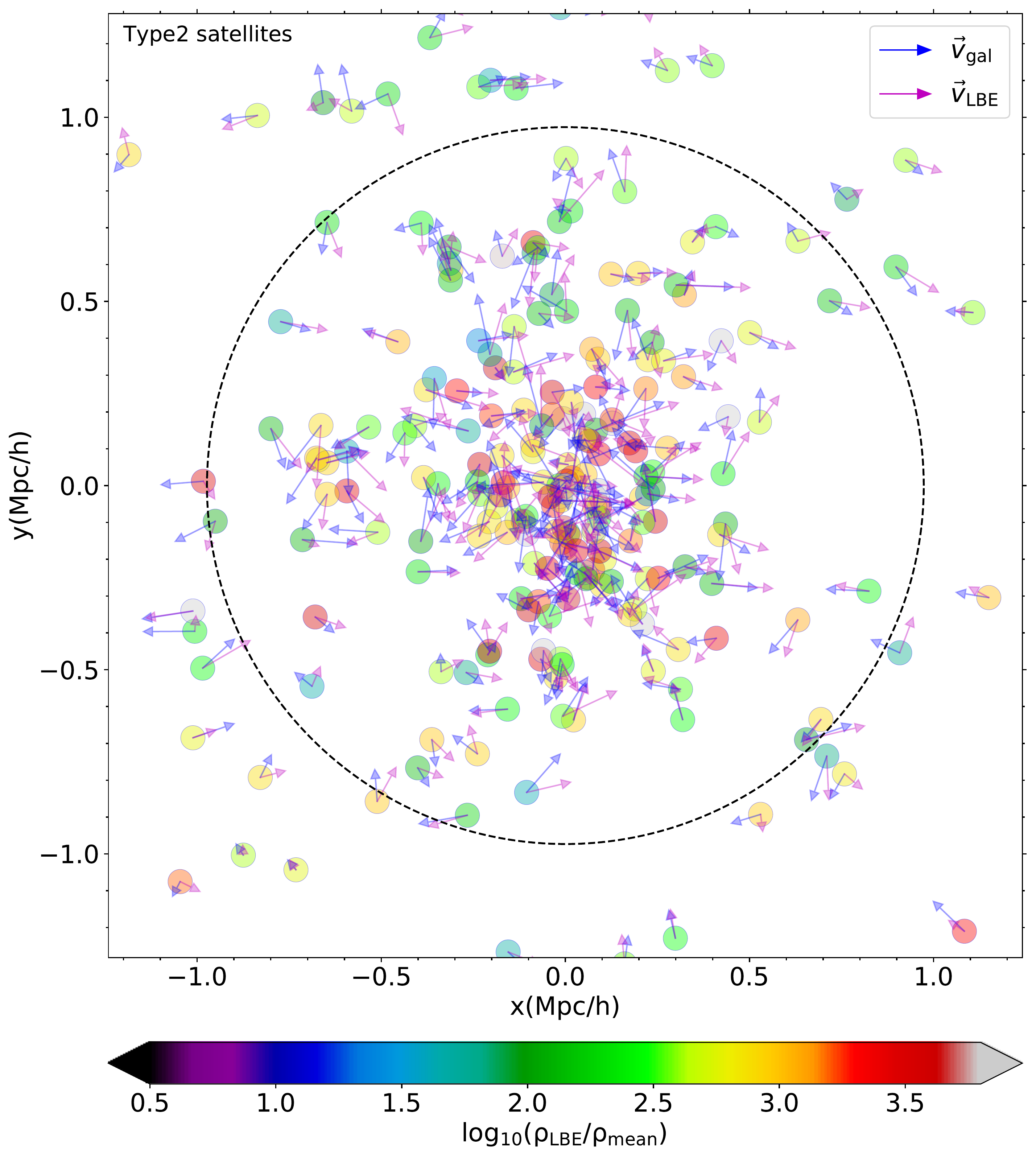}
  \caption{This figure has the same format as Fig. \ref{Fig: BGEnv_Schematic}, but is meant to contrast the locations and properties of type 1 vs type 2 in the same massive halo, zoomed into smaller scales of $\sim 1.2R_{\rm 200}$. The left panel shows type 1 satellite galaxies and the right panel shows type 2 satellites. We note that the stellar mass limit, $M_\star = 3 \times 10^9 \rm M_{\odot}/h$, is not applied in this figure. The velocity arrow scale is $0.1 {\rm Mpc/h}\equiv 1200 {\rm km/s}$. Comparing the two panels, type 2 galaxies mostly populate in the innermost regions, near the center of the main halo, and typically have higher background densities, in contrast to their type 1 counterparts.}
\label{Fig: BGEnv_Schematic21}
\end{figure*}

Fig. \ref{Fig: BGEnv_Schematic} illustrates the rich output of the LBE method for the galaxies in the vicinity of a single massive halo. Environmental properties are shown for all of the galaxies within $3\times R_{\rm 200}$ of a halo with $M_{\rm 200}\approx 3\times 10^{\rm 14} \rm M_{\odot}/h$ at redshift zero in the (scaled) Millennium Simulation. Each circle shows one galaxy, where the circle size equals the subhalo size for central (black edge colour) and type 1 satellite galaxies (red edge colour). Type 2 satellites (no edge colour) are simply shown as dots. The dashed black circle around the center corresponds to $R_{\rm 200}$ of the main halo. The arrows illustrate the velocities for central and type 1 satellite galaxies; the length of each arrow is proportional to the magnitude of the velocity. Blue arrows are the galaxy velocities, and the magenta arrows are the LBE velocities (both in the rest frame of the main halo). 

Fig. \ref{Fig: BGEnv_Schematic21} contrasts the locations and properties of type 1 vs type 2 satellites in the same halo, zoomed into a smaller scale of $1.2R_{\rm 200}$. The left panel shows type 1 satellite galaxies and the right panel shows type 2 satellites.

We point out several features of interest in Figs. \ref{Fig: BGEnv_Schematic} and \ref{Fig: BGEnv_Schematic21}. First, the LBE density generally decreases with distance from the halo center, although the lack of radial symmetry is clear. In addition, the galaxies move faster than their LBE -- for each galaxy, the blue arrow is usually longer than the magenta one. More importantly, the LBE is not completely at rest relative to its host. We also note that $\vec{v}_{\rm gal}$ and $\vec{v}_{\rm LBE}$ are more aligned for galaxies that are far away from the halo center. Near the halo center, the angle between them is statistically more uniform.  Furthermore, galaxies in high density regions move faster with respect to their local background environment. We will quantitatively explore the features mentioned here in more detail in \S \ref{sec: analyse_lbe}.

\subsection{`Decontaminated' subhalo mass function}

\begin{figure}
\centering
  \includegraphics[width=0.95\columnwidth]{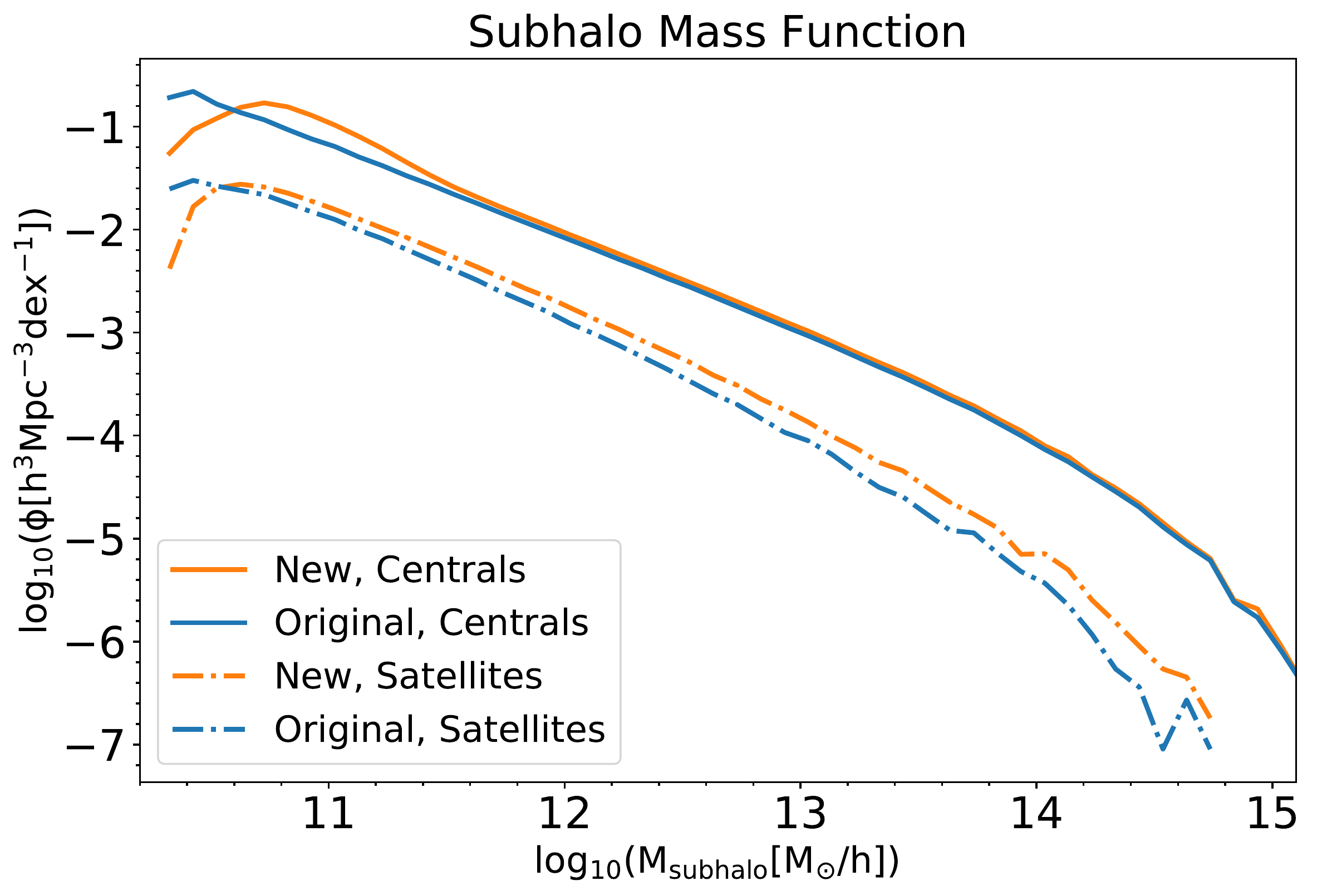}
  \includegraphics[width=0.95\columnwidth]{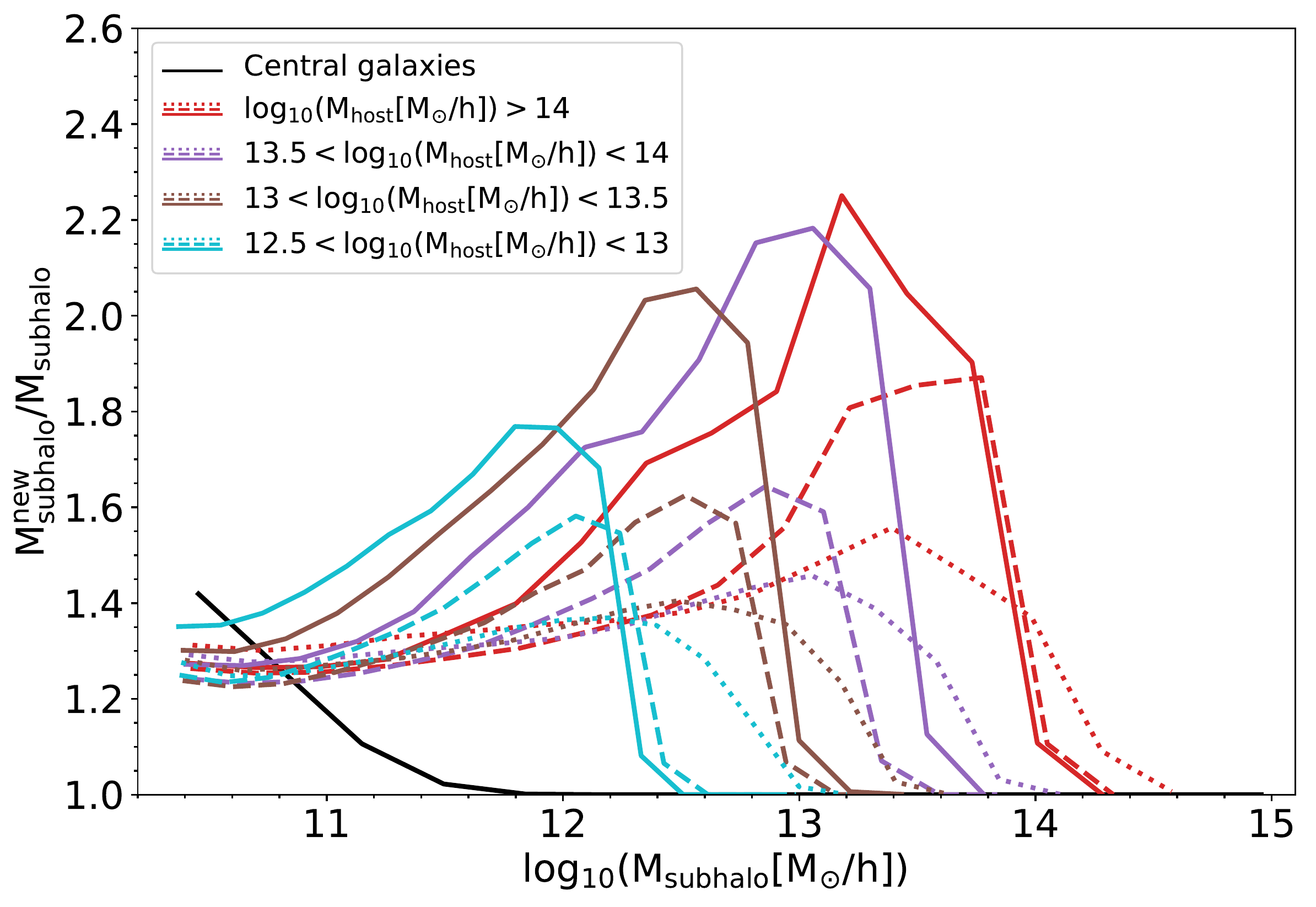}
  \caption{Top panel: comparison of the subhalo mass function estimated using the \textsc{Subfind} algorithm (blue) versus our modified mass function (orange). Solid lines include only central subhaloes, while dash-dot lines show satellites. Bottom panel: the median values of the ratio of updated versus original subhalo masses as a function of the original masses. Solid coloured lines correspond to the satellites within $0.5R_{200}$ of their FOF halo, dashed lines include $0.5<R/R_{\rm 200,host}<1$, and dotted lines show satellites beyond $R_{200}$. No mass limits are applied.}
\label{Fig: sub_massfunc}
\end{figure}

Our decontamination method (discussed in \S \ref{subsec: Gaussian_mixture}) changes the estimate of the subhalo mass, by assigning additional particles beyond the fiducial membership association determined by \textsc{Subfind}. To assess the magnitude of this change, we set the inner radius of the background shell to $R_{\rm in} = R_{\rm subhalo}$, to capture the most possible contamination, and keep the outer radius unchanged. We then add the fraction of subhalo particles in the background shell of each subhalo to its mass:

\begin{equation}
M_{\rm subhalo}^{\rm new} = M_{\rm subhalo} + f_{\rm sub}M_{\rm shell},
\end{equation}

\noindent where $M_{\rm shell}$ is the mass of background shell and $f_{\rm sub}$ is the fraction of particles in the background shell which belong to the subhalo. We show the correspondingly updated subhalo mass function for the Millennium simulation for central and satellite subhaloes in the top panel of Fig. \ref{Fig: sub_massfunc}. The orange lines show our new subhalo mass function and blue lines show the original results of \textsc{Subfind}. The solid and dashed lines represent central and satellite subhaloes. The mass function of satellite subhaloes is more affected than centrals. The bottom panel of Fig. \ref{Fig: sub_massfunc} illustrates the ratio of new to old mass for central subhaloes (black line) and satellites within FOF haloes of different masses (coloured lines).

Our results show that the subhalo mass increases somewhat for low mass centrals, $\log_{10}(M_{\rm subhalo}[ \rm M_{\odot}/h])<12$, but remains unchanged for more massive centrals. For satellite subhaloes, the changes are much larger, and increasing with increasing subhalo mass. For some subhaloes, the mass can increase by more than a factor of two. The bottom panel of Fig. \ref{Fig: sub_massfunc} shows that, at fixed subhalo mass, the mass increase is larger in lower mass FOF haloes. The bottom panel also shows that the increase in mass is larger for satellite galaxies near the center of FOF haloes (solid lines) and smaller for the ones farther away from the center (dashed and dotted lines).

We note that the masses of the most massive satellite subhaloes (the right tail of coloured curves) increase by a small fraction if at all. These are satellite subhaloes which are massive enough, compared to their host FOF, to be considered as centrals. The subhalo finder algorithm labels them as satellites because, by definition, each FOF halo can only have one central subhalo. By construction, a satellite subhalo can also not exceed the mass of its central. It is thus natural that the change in subhalo mass for these objects drops to the same value as for central subhaloes, even though they are categorized as satellites. 


\section{Analysis of LBE properties}
\label{sec: analyse_lbe}

In this section, we analyze the statistical properties of the LBE. We look for correlations between the LBE and subhalo properties including mass, type (central or satellite),  position within the parent FOF halo, and mass of the host FOF halo (for satellites). This provides a first hint of the importance of ram-pressure stripping as a function of these same properties. 

\subsection{Correlation of LBE with subhalo mass}

\begin{figure}
\centering
  \includegraphics[width=0.95\columnwidth]{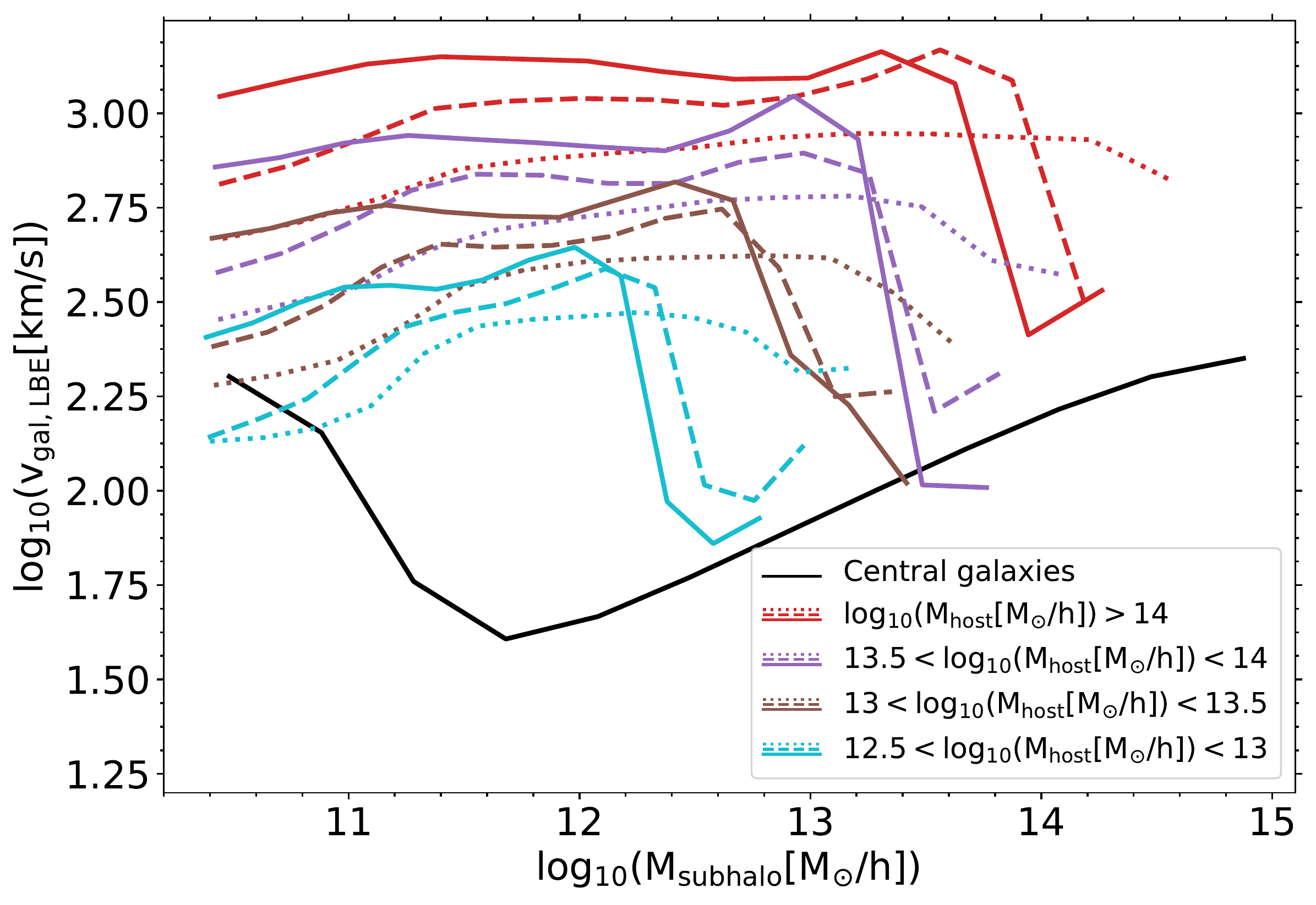}
  \includegraphics[width=0.95\columnwidth]{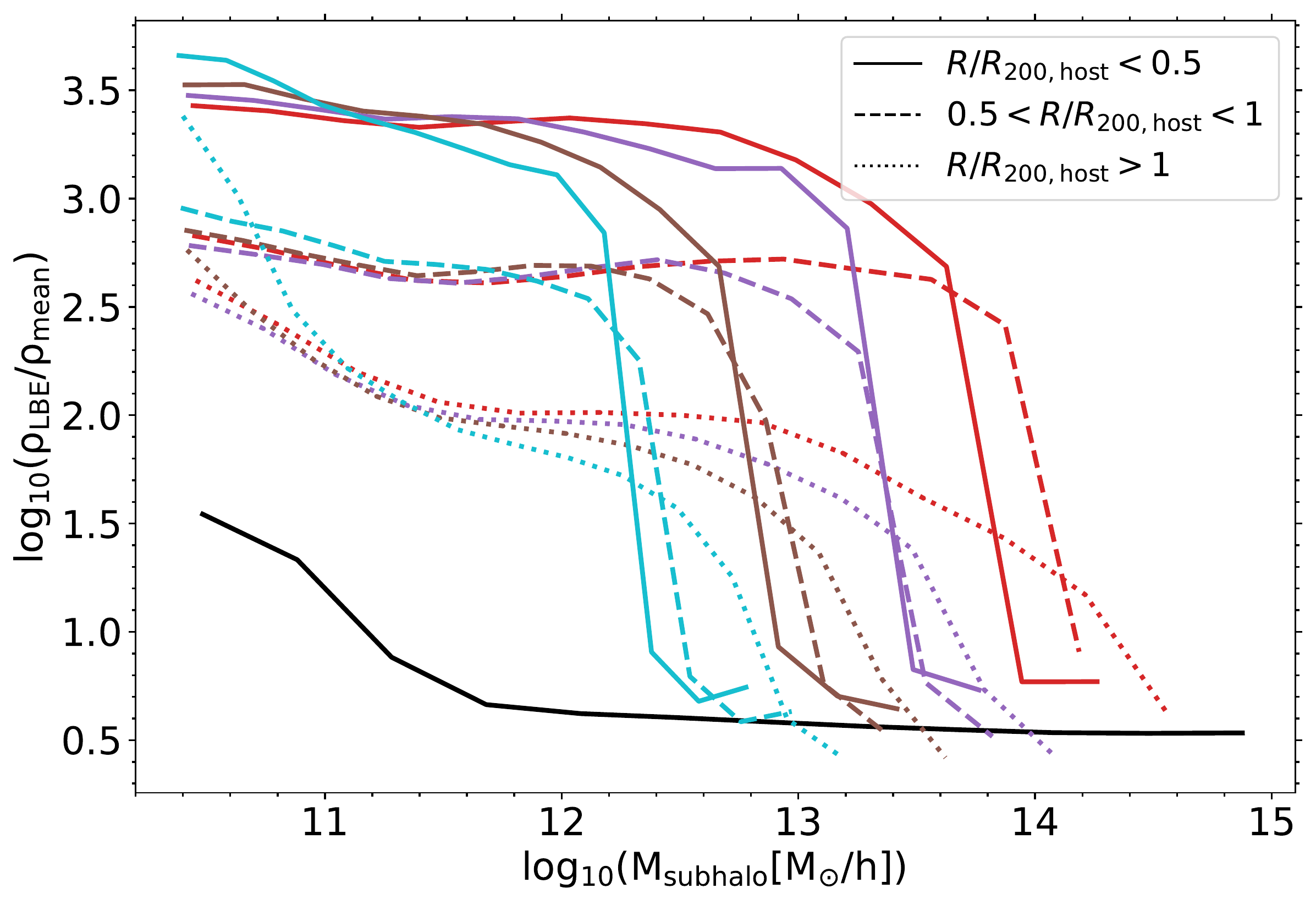}
  \includegraphics[width=0.95\columnwidth]{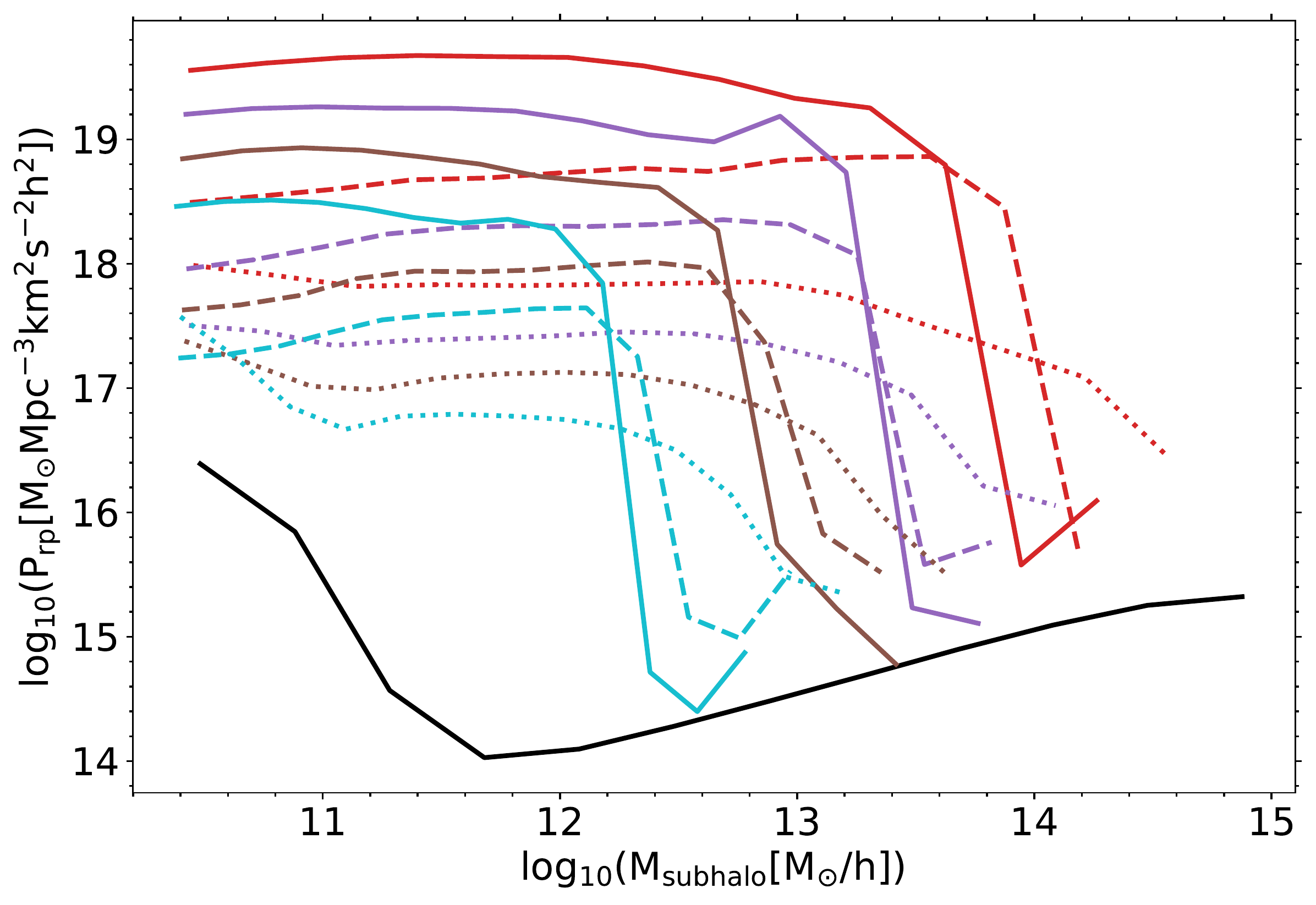}
  \caption{The velocity of galaxies relative to their LBE (top panel), the LBE density (middle panel) and ram-pressure force (bottom panel) as a function of subhalo mass for all of the subhaloes at $z=0$. The black solid line corresponds to central galaxies, while the coloured lines show satellites. Solid coloured lines include satellites within $0.5 R_{200}$ of the centre of the FOF halo, dashed lines are for $0.5<R/R_{\rm 200,host}<1$ and the dotted lines show satellites beyond $R_{200}$. }
\label{Fig: Median_BGEnv}
\end{figure}

Fig. \ref{Fig: Median_BGEnv} depicts galaxy velocity with respect to the local environment, $v_{\rm gal,LBE}$ (top panel), density $\rho_{\rm LBE}$ (middle panel) and ram-pressure force $P_{\rm rp}$ (bottom panel) as a function of subhalo mass for central and type 1 satellite galaxies at $z=0$ in the Millennium Simulation. In all the panels, black solid lines show medians for central galaxies, and coloured lines given medians for satellite galaxies, where different colours denote different host halo masses: within $R_{\rm 200} / 2$ (solid), with $0.5 < R/R_{\rm 200} < 1$ (dashed), and beyond $R_{\rm 200}$ (dotted).

For satellite galaxies, the trends of all the three physical quantities have a weak dependence on subhalo mass up to a threshold mass, beyond which there is a sharp drop. This drop happens, as before, when the subhalo mass approaches that of the central subhalo, and these massive subhaloes become equivalent to centrals in terms of their properties. Binary subhaloes at the center of a FOF halo are expected to merge, but before they do so, one subhalo will be classified as a central and the other as a satellite of much lower mass.

We also see that $\rho_{\rm LBE}$ has a weak dependency on the host mass for low mass satellite galaxies, whereas $v_{\rm gal,LBE}$ increases strongly with host mass, indicating that satellites of more massive haloes move faster relative to their LBE. The same trend is seen for the ram-pressure since it is proportional to $v_{\rm gal,LBE}^2$. Indeed, the median ram-pressure force is a factor of 10 larger for satellites within massive clusters of $10^{14} M_{\odot}$ than for satellites within groups of $10^{12}-10^{13} M_{\odot}$. There is also a significant dependence of all three quantities on distance from the center of the host halo. We note that this distance dependence is much stronger for $\rho_{\rm LBE}$ than for $v_{\rm gal,LBE}^2$. Moreover, for $v_{\rm gal,LBE}$, this trend is stronger in more massive haloes.

\subsection{LBE properties versus distance}
\label{subsec: LBE_dis}

As seen in the previous section, the LBE in the vicinity of satellite galaxies changes with the distance from the centre of the FOF halo as well as with the mass of the FOF halo. To investigate this in more detail, we extend our study out to $5R_{\rm 200}$ around each FOF halo and consider all galaxies contained in that volume. 

\begin{figure}
\centering
\includegraphics[width=0.95\columnwidth]{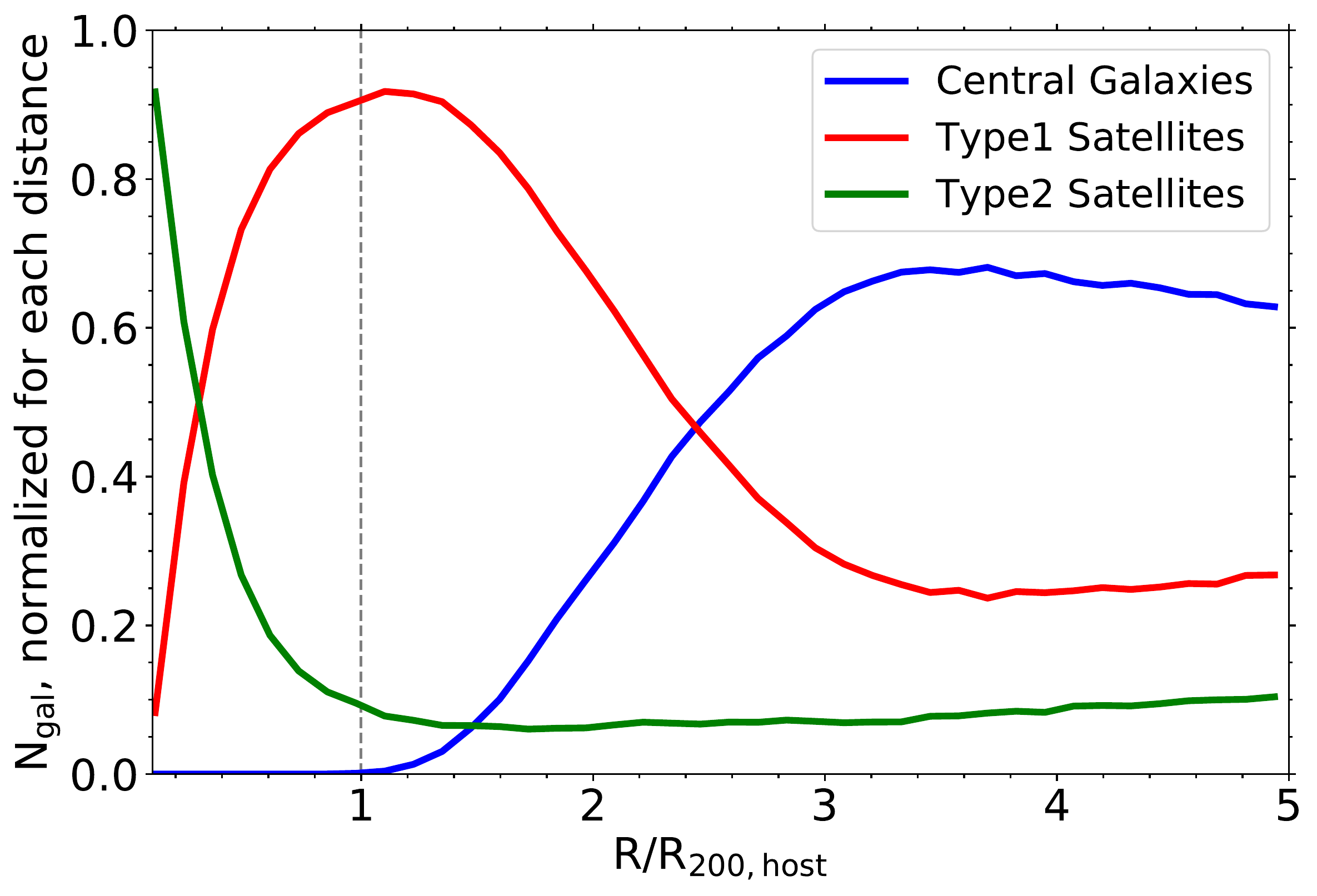}
\caption{Distribution of different types of galaxies in \textsc{L-Galaxies} as a function of distance from the centre of FOF haloes. Blue lines correspond to central galaxies and red (green) lines to type 1 (type 2) satellite galaxies. At each distance, the sum of the three curves is normalized to unity.}
\label{fig: types_dis_pdf}
\end{figure}

Fig. \ref{fig: types_dis_pdf} shows the distribution of the three different galaxy types as a function of their distance from the center of their FOF halo. The curves are normalized to the total number of galaxies in each distance bin, such that the sum of three curves is equal to unity. It can be seen that the type 2 satellites dominate close to the center of the FOF halo, while beyond $R_{\rm 200}$ they form about 10\% of the population. In contrast, type 1 satellites are most frequent at radii $0.5<R/R_{\rm 200}<2.5$. Central galaxies start to be dominant from $\approx 2.5R_{\rm 200}$. Around $R> 3.5R_{\rm 200}$ the distributions no longer vary with  distance. At these large radii, there are approximately 65\% centrals, $\sim$ 25\% type1 satellites and $\sim$10\% type 2 satellite galaxies, and these fractions are insensitive to central halo mass. 

\begin{figure*}
  \includegraphics[width=0.9\textwidth]{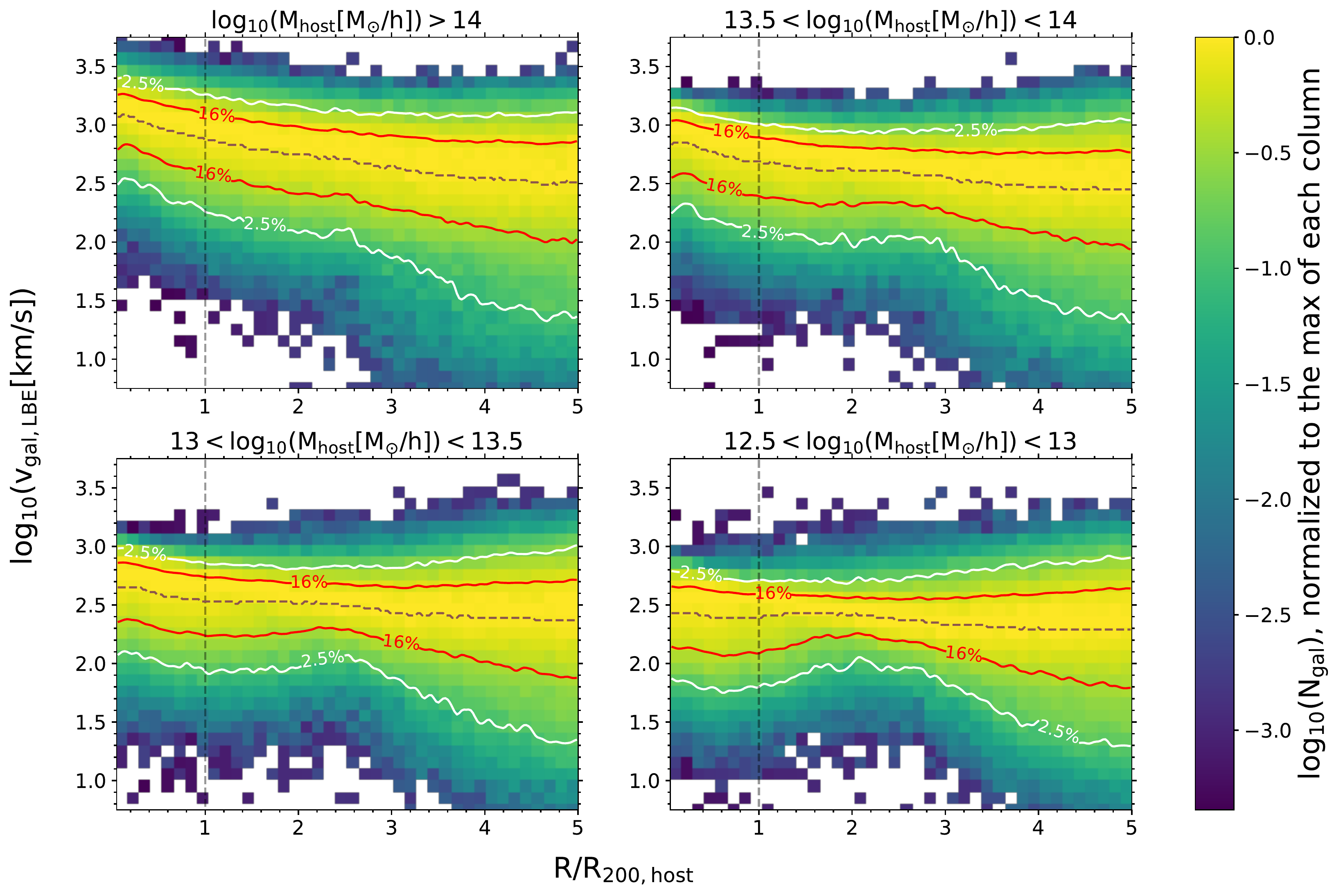}
    \includegraphics[width=0.9\textwidth]{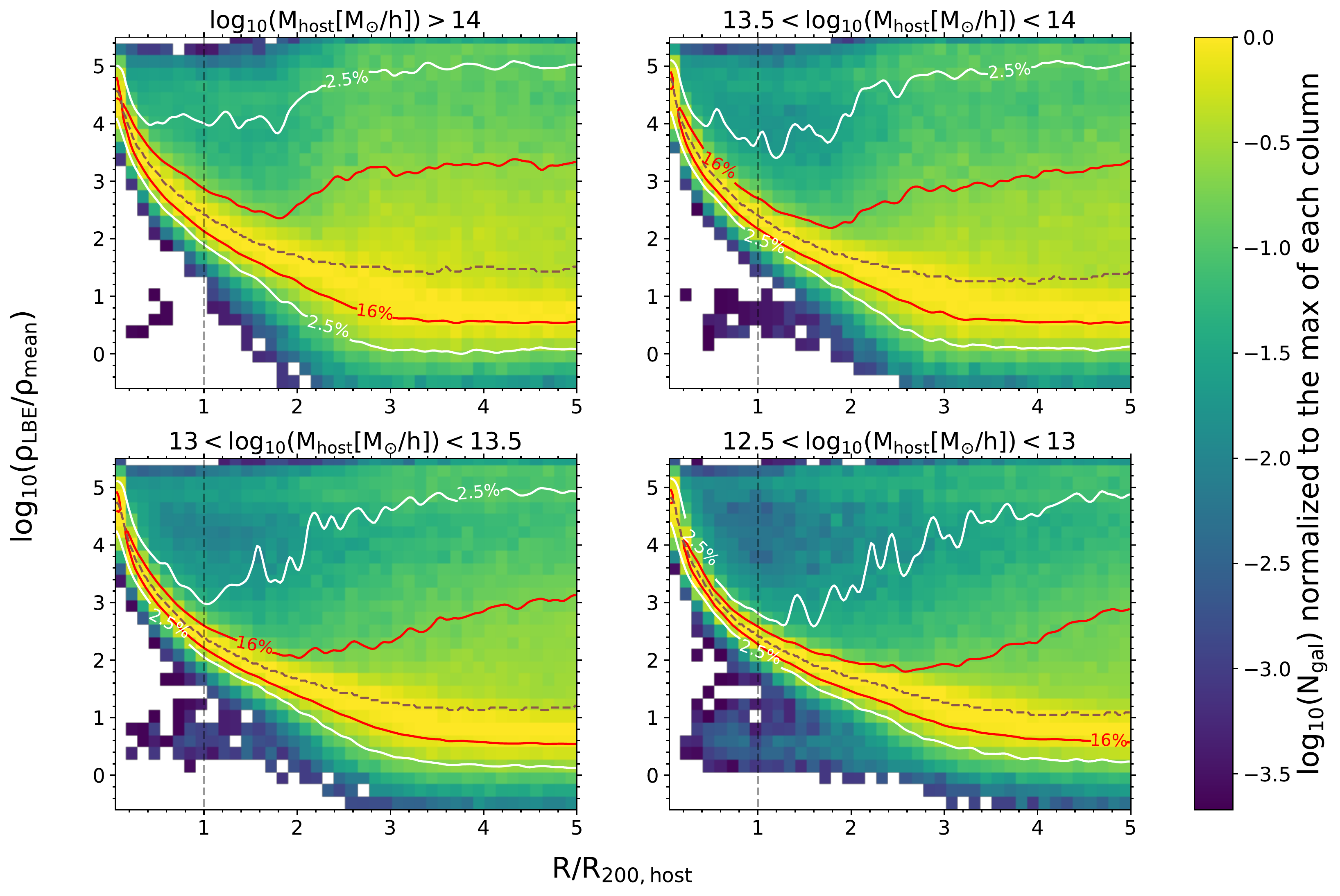}
  \caption{Distribution of galaxy velocities relative to their LBE (top 4 panels) and LBE density (bottom 4 panels) as a function of distance from the centre of the host halo. Each group of 4 panels shows results for four different FOF host halo mass bins, and we include all galaxies out to $5R_{\rm 200,host}$. The colour table is logarithmic and values at each distance column are normalized to the maximum value of that column. Red and white contours encompass the $1\sigma (\approx 68\%)$ and $2\sigma (\approx 95\%)$ variation in properties at each distance. The brown dashed line shows the median value at each distance and the vertical gray dashed lines mark the virial radius.}
\label{Fig: VelDisPDF}
\end{figure*}

We continue by investigating how $v_{\rm gal,LBE}$ and $\rho_{\rm LBE}$ change as a function of distance from the center of the FOF halo. The results are shown in four bins of FOF halo mass $M_{200}$ in Fig. \ref{Fig: VelDisPDF}, which illustrates $v_{\rm gal,LBE}$ (top 4 panels) and $\rho_{\rm gal,LBE}$ (bottom 4 panels) as a function of distance to the centre of the FOF haloes. We use bins of width $0.1/R_{\rm 200,host}$ in distance, and normalize all the pixels at each distance to the maximum value at that distance. In addition, the $1\sigma$ (68\%) and $2\sigma$ (95\%) points of the galaxy distribution at each radius are indicated by red and white contours, while the brown dashed lines show the median value. We combine all three galaxy types together in this and the following figures.

Comparing the top four panels in Fig. \ref{Fig: VelDisPDF}, we see that, at small radii, the median value of $v_{\rm gal,LBE}$ increases with the host mass, reflecting the fact that velocity dispersion is higher in more massive haloes. We recall that $v_{\rm gal,LBE}$ is the velocity of a galaxy relative to its local background environment. Comparing the bottom four panels, we see that the median background density shows a weaker dependence on the host mass. However, the distribution of density exhibits more scatter at large distances, particularly in the vicinity of massive compared to low mass haloes. Distant galaxies with high LBE densities are mostly satellites (of other parent haloes). At $R/R_{\rm 200,host} = 5$, the background density around satellite galaxies is $\sim$ 2 dex larger than around centrals at the same distance. At large distances $v_{\rm gal,LBE}$ is also considerably different for central and satellite galaxies. Central galaxies move a factor of two slower relative to their LBE compared to satellites at a distance of $5R_{200}$.

Interestingly, from Fig. \ref{Fig: VelDisPDF} it can be seen that both $v_{\rm gal,LBE}$ and $\rho_{\rm LBE}$ show quite continuous behavior at all scales, including across $R_{200}$ (dashed black line). This implies that there is no abrupt truncation of environmental effects such as ram-pressure stripping at the virial radius, at least in the population average sense. 

\begin{figure*}
  \includegraphics[width=0.9\textwidth]{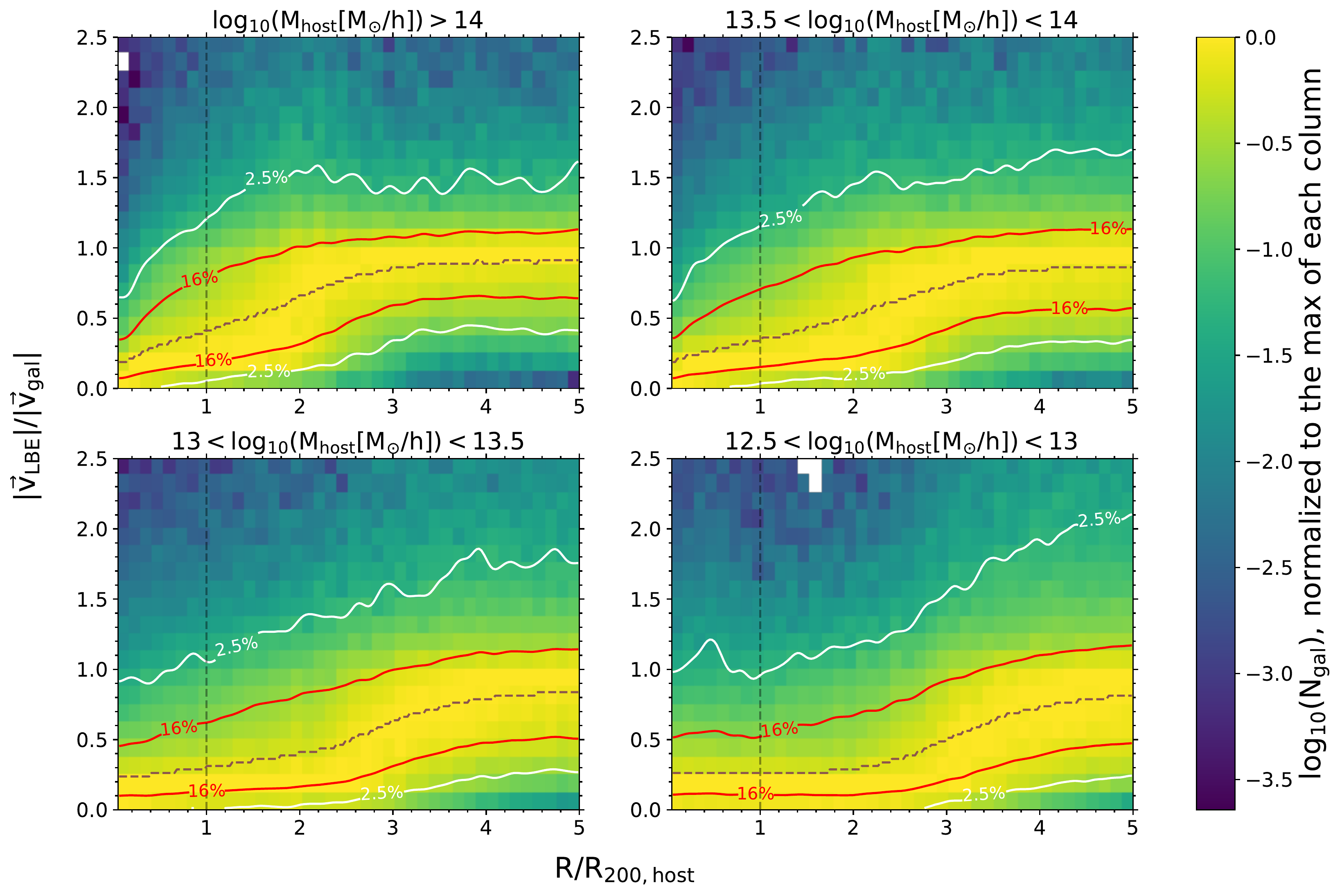}
  \includegraphics[width=0.9\textwidth]{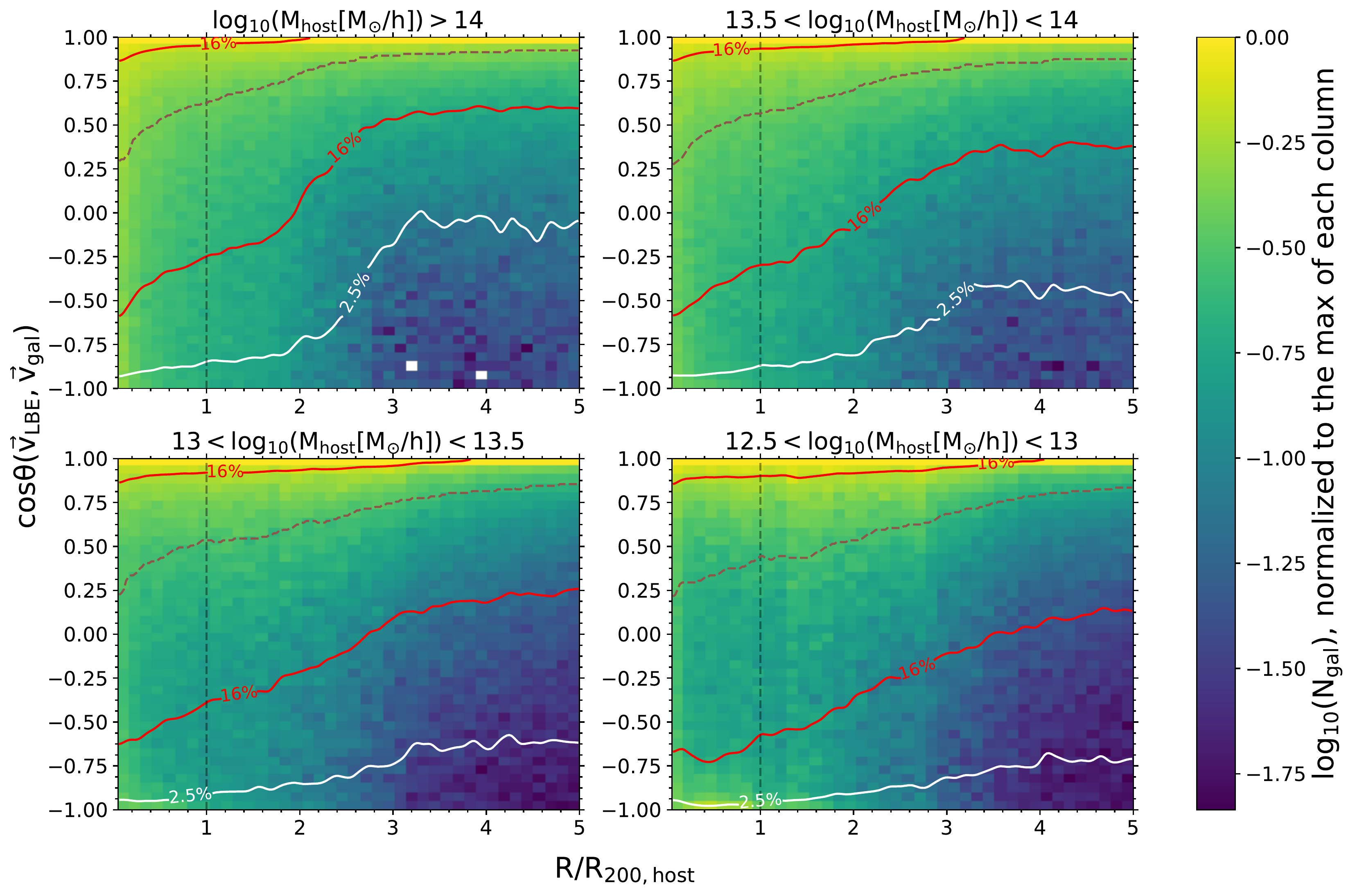}
  \caption{Distribution of the ratio (top panels) and angle (bottom panels) between LBE velocity and galaxy velocity (both in the rest-frame of central host halo) as a function of distance from the centre of the host halo. The format is the same as in Figure \ref{Fig: VelDisPDF}.}
 \label{Fig: VBgOVsub}
\end{figure*}

We now turn to a comparison of the motion of galaxies with respect to the motion of their LBEs. Fig. \ref{Fig: VBgOVsub} shows how $\vec{v}_{\rm LBE}$ and $\vec{v}_{\rm gal}$ differ in magnitude and orientation. The colours and contours are the same as in Fig. \ref{Fig: VelDisPDF}. The top panels demonstrate that the LBE velocity is on average larger than $v_{\rm gal}/4$ in the centers of haloes. It is clear that the common assumption of a LBE for satellite galaxies that is at rest with respect to the hosts \cite[e.g.][]{stevens2016building,simpson2018quenching} does not hold in general.

\begin{figure*}
  \includegraphics[width=0.9\textwidth]{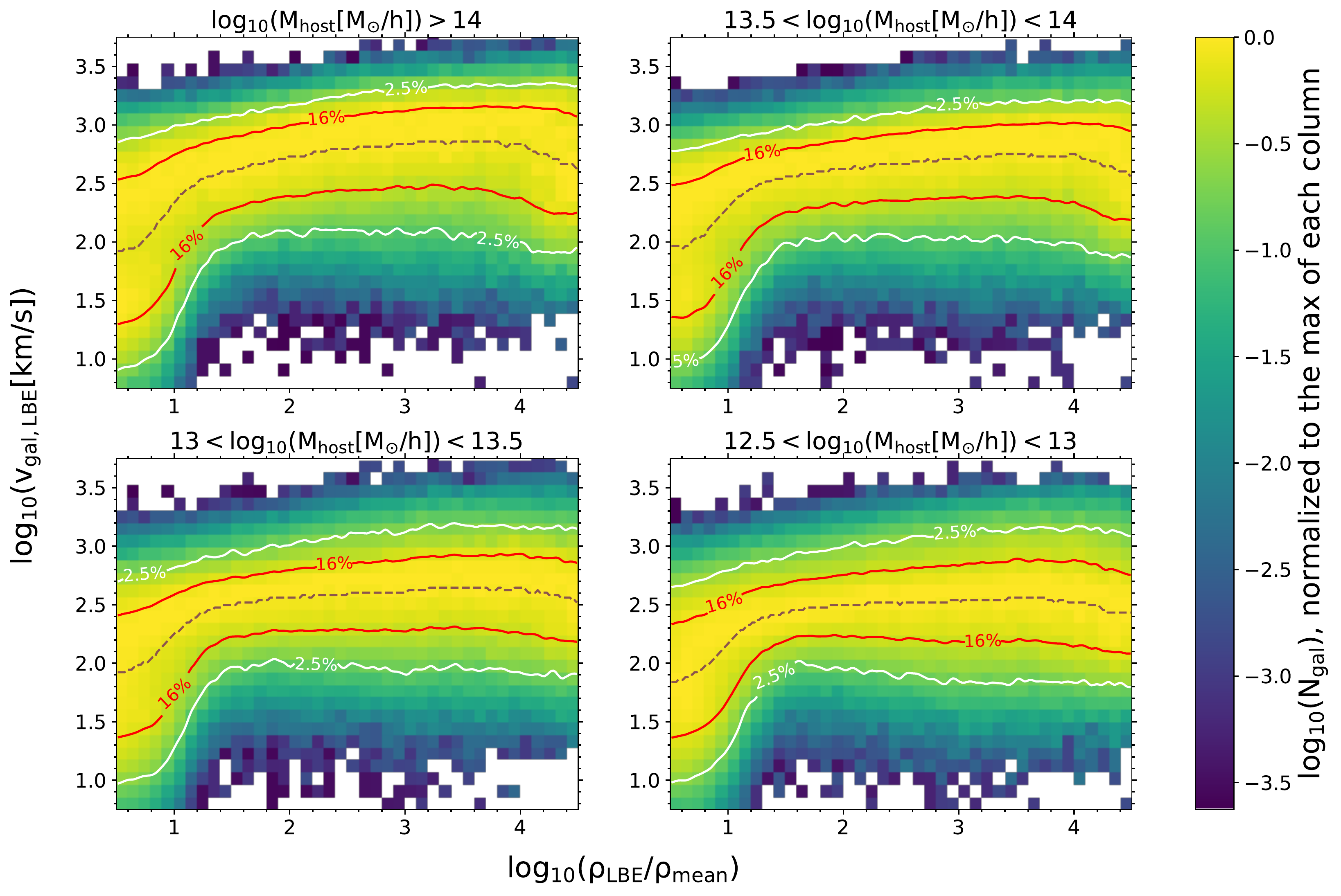}
  \caption{Distribution of the velocity of galaxies relative to their LBE as a function of their LBE density. The format of the panels is the same as in Figure \ref{Fig: VelDisPDF}. To first order density traces distance, and only in low-density regimes $\log(\rho_{\rm LBE}/\rho_{\rm mean}) \lessapprox 1.5$ does $v_{\rm gal,LBE}$ increase strongly with background density. This threshold overdensity roughly corresponds approximately to galaxies on first infall.}
\label{Fig: VelRhoPDF}
\end{figure*}

In all four mass bins, the velocity of the LBE at large distance asymptotically approaches a characteristic value of $\sim$ 80\% of the galaxy velocity itself. This value is reached only at progressively larger distances when normalizing by $R_{\rm 200}$, for lower mass haloes. Near massive haloes, we interpret this as a signature of coherent infall, with galaxies and their surrounding dark matter accreting at similar velocities.

The four bottom panels in Fig. \ref{Fig: VBgOVsub} show the angle between $\vec{v}_{\rm gal}$ and $\vec{v}_{\rm LBE}$, both in the rest frame of the central FOF halo. At larger distances ($R \gtrapprox R_{\rm 200,host}$), the two noted velocities are frequently aligned in the same direction -- as explained above, both the galaxy and its surrounding material move coherently in the same direction towards a nearby massive halo or mass concentration. This large-scale alignment is more pronounced around more massive centrals, as expected.

The distribution of angles becomes more uniform as $R\rightarrow0$, reflecting the fact that the subhalo orbits become isotropized close to the center of haloes. The median stacked cosine of the relative angle drops below $1/2$ roughly within $R_{\rm 200}$. Even at the halo center, however, the $\cos(\theta)$ distributions do not become entirely symmetric about zero, showing that the dynamical states of groups and clusters are in general not fully relaxed.

\subsection{Galaxy velocity versus density}

Finally, we examine how the galaxy's speed relative to its LBE changes with LBE density. Fig. \ref{Fig: VelRhoPDF} shows $v_{\rm gal,LBE}$ as a function of $\rho_{\rm LBE}$, for all galaxies within $5 R_{\rm 200}$, stacking around central objects in the same four mass bins as before. At the centers of haloes, roughly independent of host mass, there is a residual $\sim$ 100 km/s motion between galaxies and their local background.

We find that the median relative velocity of galaxies with respect to their LBE increases steeply with LBE density in low density regions, i.e. $\log(\rho_{\rm LBE}/\rho_{\rm mean}) \lessapprox 1.5$. In higher density regions, on the other hand, $v_{\rm gal,LBE}$ is only weakly dependent on LBE density. However, at these high densities, galaxy velocity shows significant correlation with FOF halo mass. We note that the density where $v_{\rm gal,LBE}$ becomes constant corresponds roughly to the density where galaxies are expected to be infalling for the first time. Notably, the imprint of local density on $v_{\rm gal,LBE}$ is much more apparent than the imprint of radius from the FOF halo center, where these two regimes were much less clearly evident.


\section{Results of \textsc{L-Galaxies} with our new stripping model}
\label{sec: SAM}

\begin{figure}
\centering
\includegraphics[width=0.95\columnwidth]{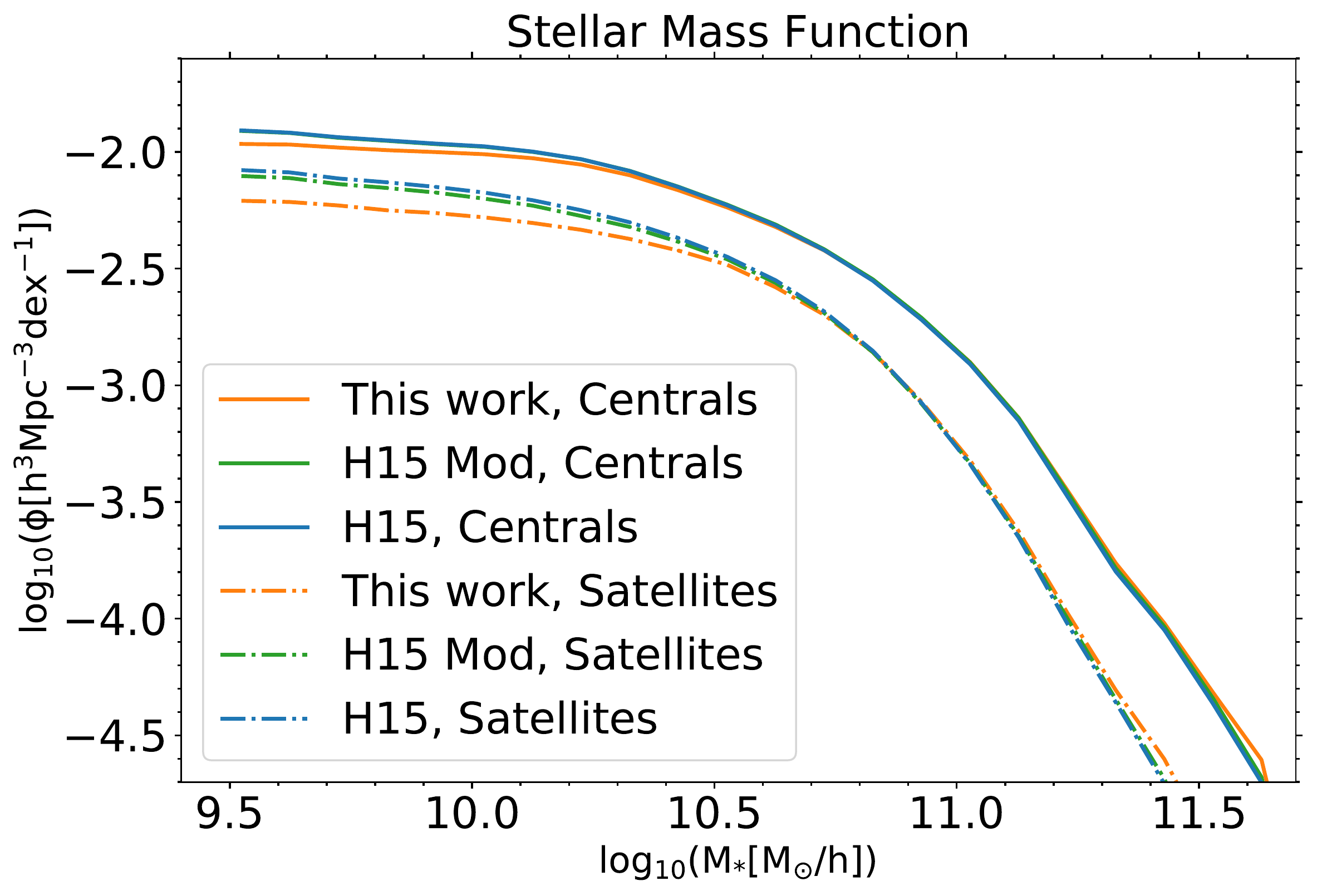}
\includegraphics[width=0.95\columnwidth]{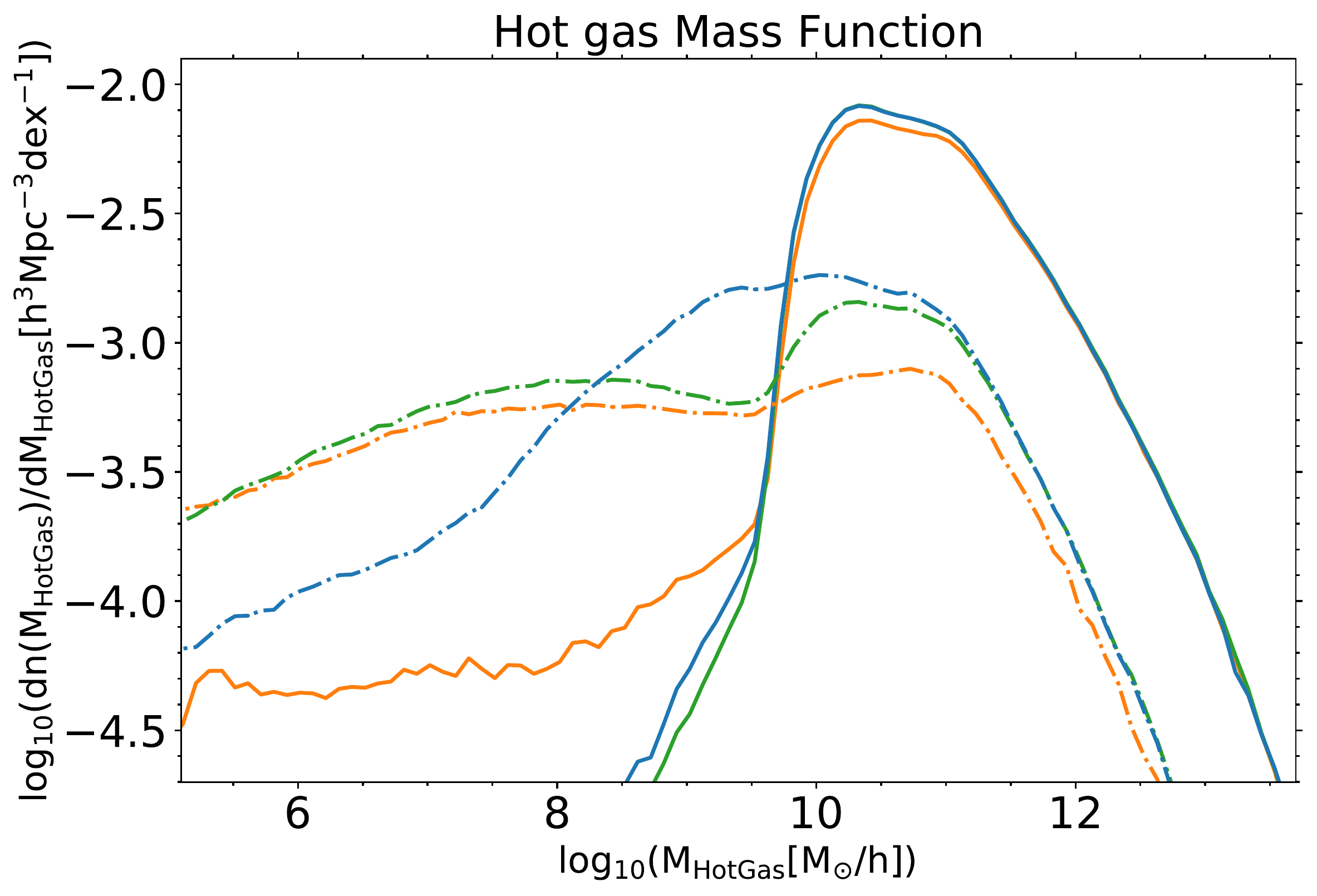}
\caption{Stellar mass function (top panel) and hot gas mass function (bottom panel) for galaxies at $z=0$. The orange line shows results from our new model, while green and blue lines show the outcomes of the H15 modified (see text for explanation) and H15 models, respectively. Centrals (solid) and satellites (dash-dot) are shown separately.}
\label{fig: Mass_Functions}
\end{figure}

The majority of galaxies in the simulation are either central galaxies or satellites which reside beyond $R_{200}$ of the more massive FOF haloes. At $z=0$ of our \textsc{L-Galaxies} run on MS, about 20\% of all galaxies are satellites within $R_{200}$ of FOF haloes. As we have shown in \S \ref{sec: analyse_lbe}, LBE properties are generally continuous across $R_{200}$, motivating our extension of RPS to all galaxies in the simulation regardless of type or location. As a result, the bulk of the galaxy population will be newly affected by stripping.

In this section, we incorporate the information provided by the LBE measurements into our new RPS model implemented in the \textsc{L-galaxies} code (\S \ref{sec: rps and tidal}). We compare our model with (i) the fiducial H15 result, and (ii) a modified version of H15 (hereafter `H15 Mod'). In H15, the RPS for hot gas is limited to satellites within $R_{200}$ of haloes more massive than $2\times10^{14}\rm M_{\odot}/h$. In H15 Mod, we remove this halo mass limit and apply the H15 RPS prescription to all satellites within $R_{200}$, regardless of host mass. 

We report our results for galaxies at $z=0$, as the properties of galaxies at redshift zero are influenced by the physical processes they have experienced through their history. This allows us to assess the integrated influence our new RPS model. 

\subsection{Amount and impact of hot gas stripping}

\subsubsection{Stellar and hot gas mass functions}
\begin{figure*}
\includegraphics[width=0.9\columnwidth]{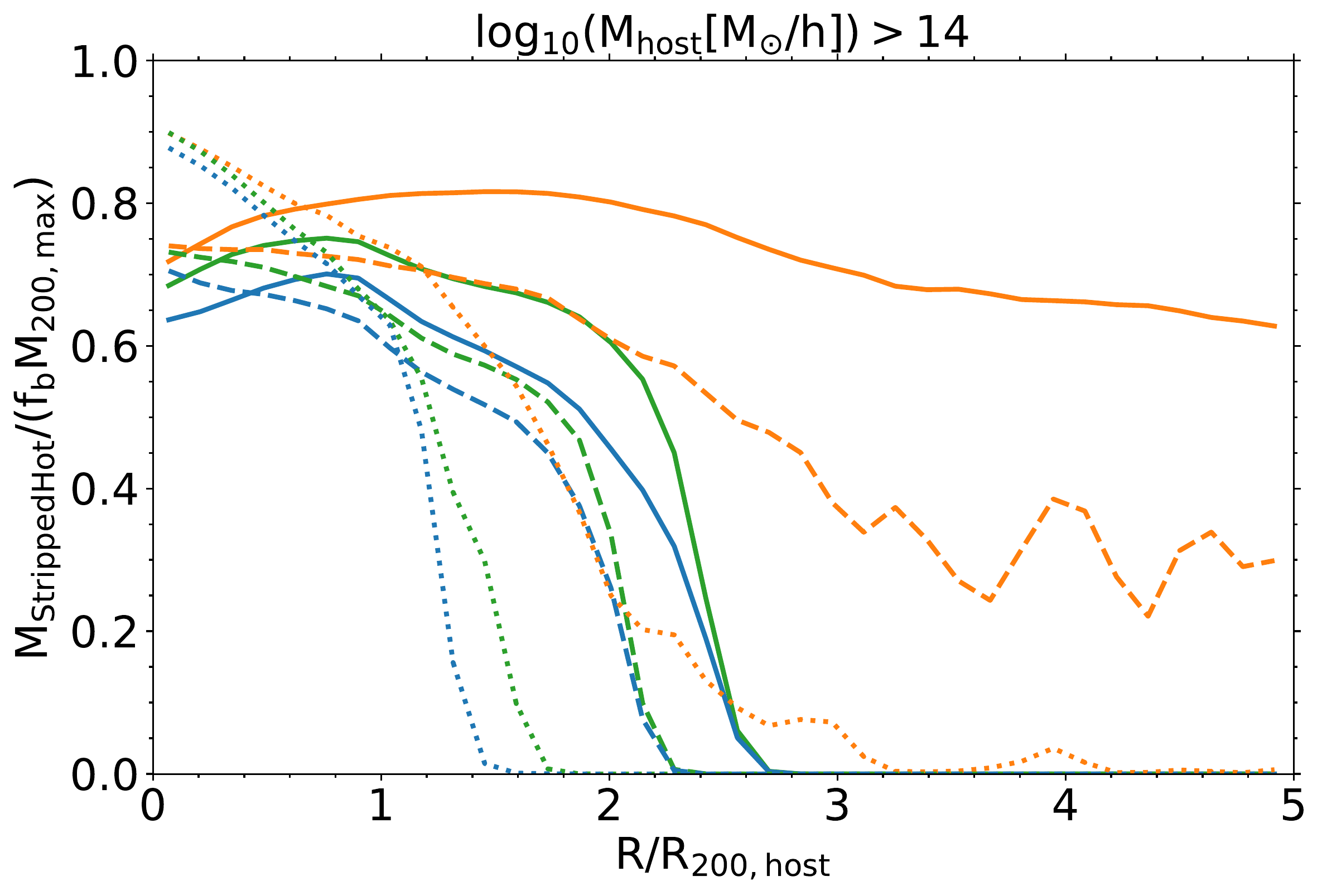}
\includegraphics[width=0.9\columnwidth]{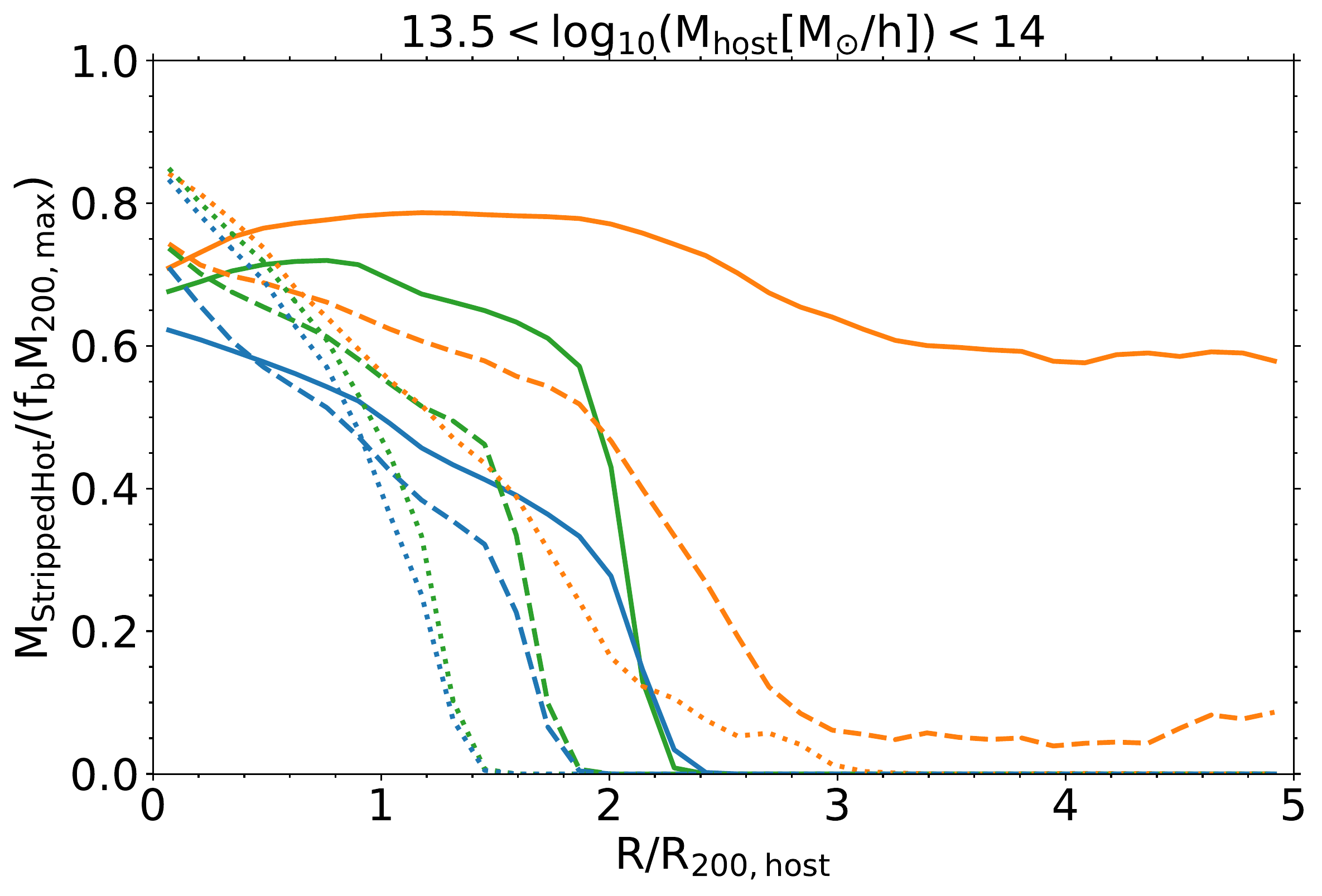}
\includegraphics[width=0.9\columnwidth]{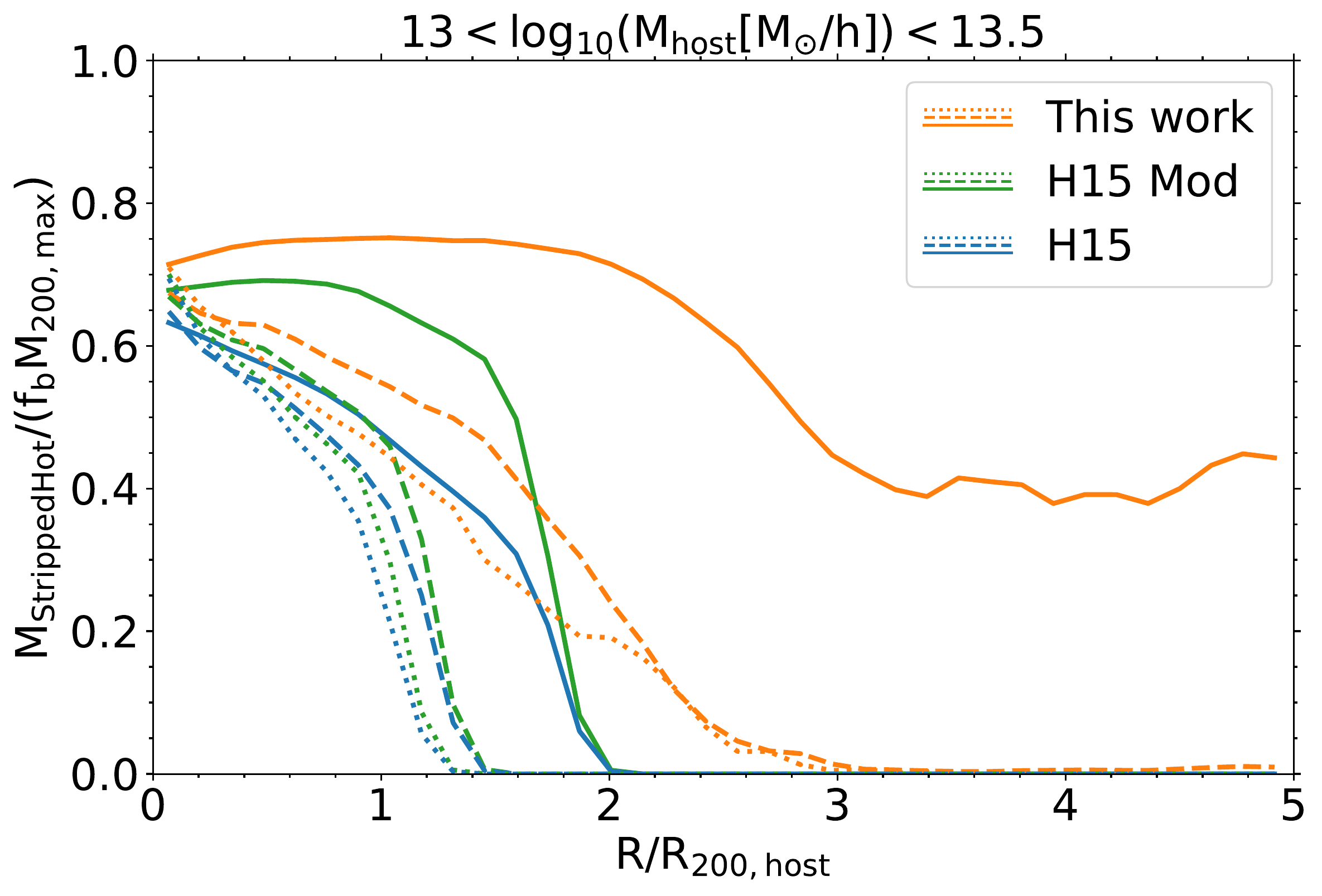}
\includegraphics[width=0.9\columnwidth]{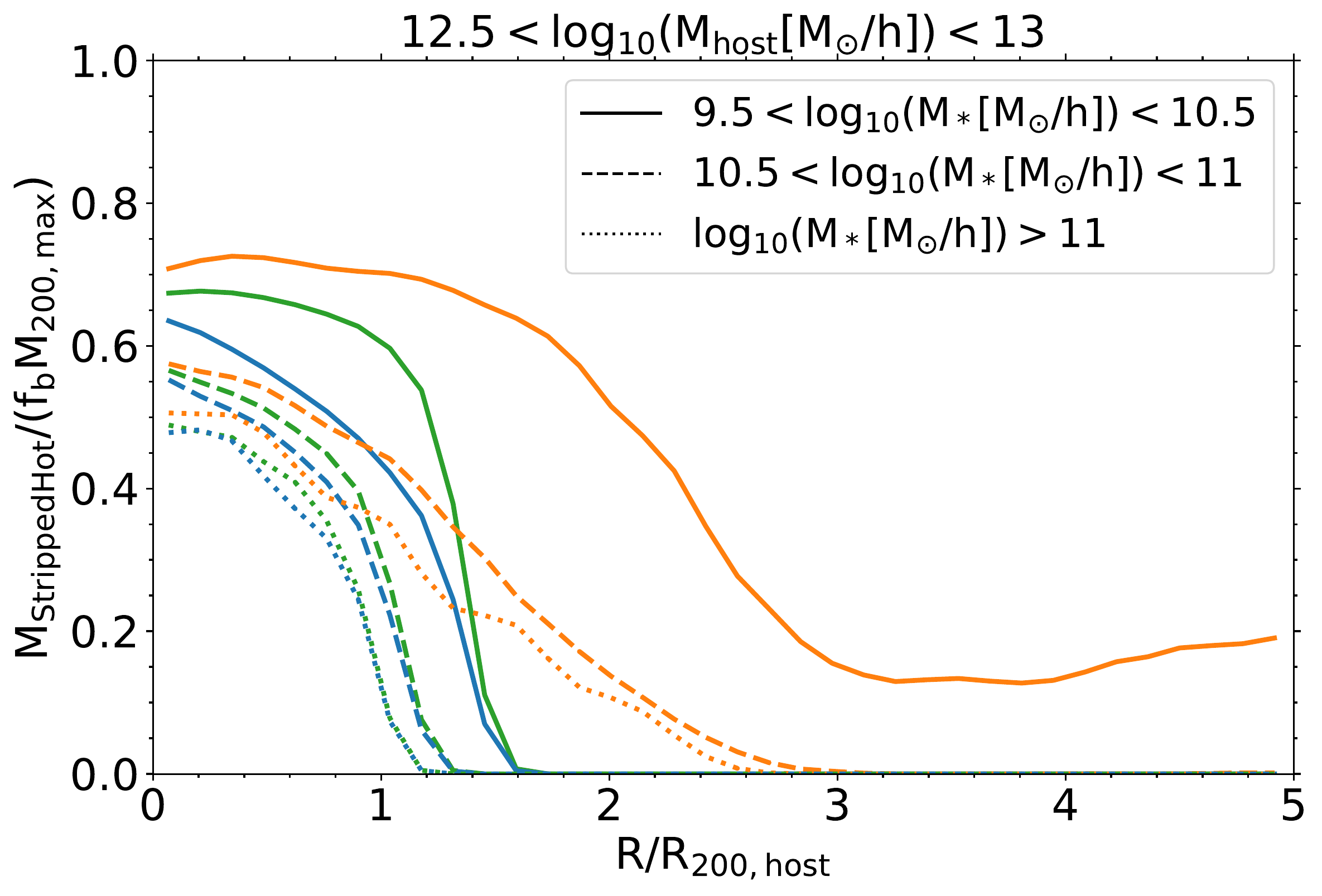}
\caption{Median stripped hot gas fraction of galaxies as a function of their distance from the centre of the  host halo. The total stripped hot gas fraction is defined as $f_{\rm StrippedHot} = M_{\rm StrippedHot}/(f_{\rm b}M_{\rm 200,max})$, where $f_{\rm b}$ is the cosmic baryon fraction and $M_{\rm 200,max}$ is the maximum past $M_{200}$. $M_{\rm StrippedHot}$ is the sum of stripped hot gas mass over all time-steps for the main progenitor of the galaxy. Different panels show results for different central host halo masses. In all of the panels, we include our new model (orange), H15 (blue), and H15 Mod (green). Different line styles correspond to different stellar mass ranges: solid lines are for galaxies with $9.5 < \log(M_\star [\rm M_{\odot}/h]) < 10.5$, dashed lines for $10.5 < \log(M_\star [\rm M_{\odot}/h]) < 11$, and dotted lines for $\log(M_\star [\rm M_{\odot}/h]) > 11$.}
\label{fig: median_totStrHot_MStellar}
\end{figure*}

Among all of the physical quantities predicted by the semi-analytic model, stellar mass is one of the least affected by our changes to RPS. This allows us to report changes to all other quantities as a function of stellar mass. In contrast, the hot gas mass function significantly changes in our model, especially for satellite galaxies. Fig. \ref{fig: Mass_Functions} illustrates the stellar (top) and hot gas (bottom) mass functions at $z=0$. Solid lines denote central galaxies and dash-dotted lines correspond to satellites.

It can be seen that the stellar mass function in our new model (orange lines) has changed only slightly. The effect is strongest for the least massive galaxies, which are more strongly influenced by RPS. At high mass, beyond the knee at $M_\star \simeq 10^{10.5} \rm{M}_\odot$, there is no appreciable change to the stellar masses of the overall galaxy population with respect to H15 (blue lines) or H15 Mod (green lines). As the SMF is already tightly constrained in H15 by observational measurements, it is reassuring that it is not modified by our new RPS treatment.

The bottom panel of Fig. \ref{fig: Mass_Functions} shows the hot gas mass functions for these three models. In contrast to $M_\star$, we observe significant changes across almost the entire range of total hot gas masses. There are now a noticeable fraction of central galaxies with low hot-gas mass, $\log(M_{\rm hotgas}[\rm M_{\odot}/h]) \lesssim 9.0$. Stripping in these central galaxies is primarily driven by rapid movement with respect to their LBE. Without RPS, they appear in the $\log(M_{\rm hotgas}[\rm M_{\odot}/h]) \gtrsim 10$ portion of the distribution. The difference at the peak of the mass function between the three models demonstrates that this is a small effect by number. For satellite galaxies, on the other hand, this effect is much more significant. Looking at the peak of the hot gas distribution in the bottom panel of Fig. \ref{fig: Mass_Functions}, the value in our new model is about 0.4 dex smaller than in previous models. In general, we find that the hot gas mass function in our new model is usually below those of H15 and H15 Mod in the mass range $\log(M_{\rm hotgas}[\rm M_{\odot}/h])\gtrsim6$ for satellite galaxies.

\begin{figure*}
\includegraphics[width=0.9\columnwidth]{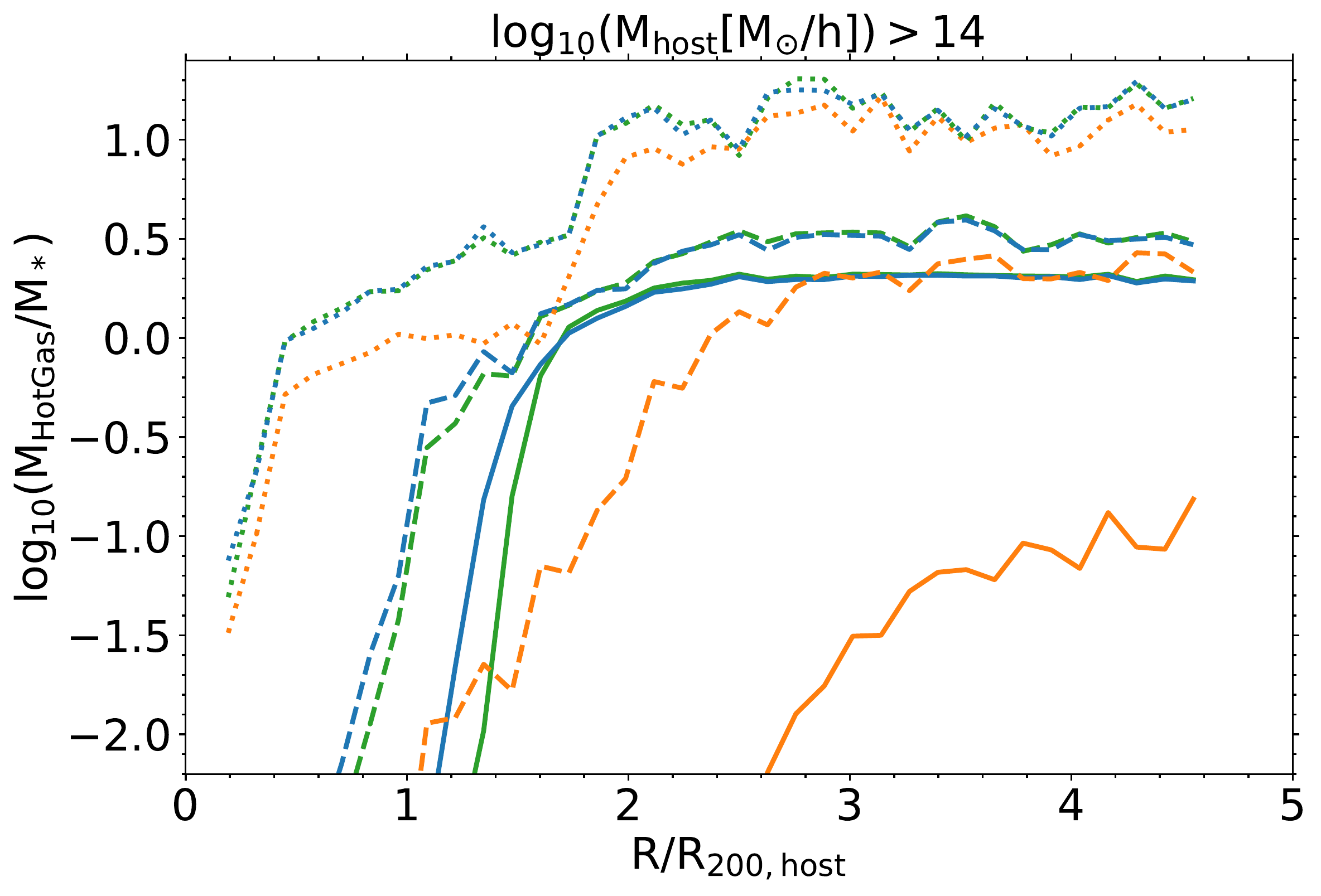}
\includegraphics[width=0.9\columnwidth]{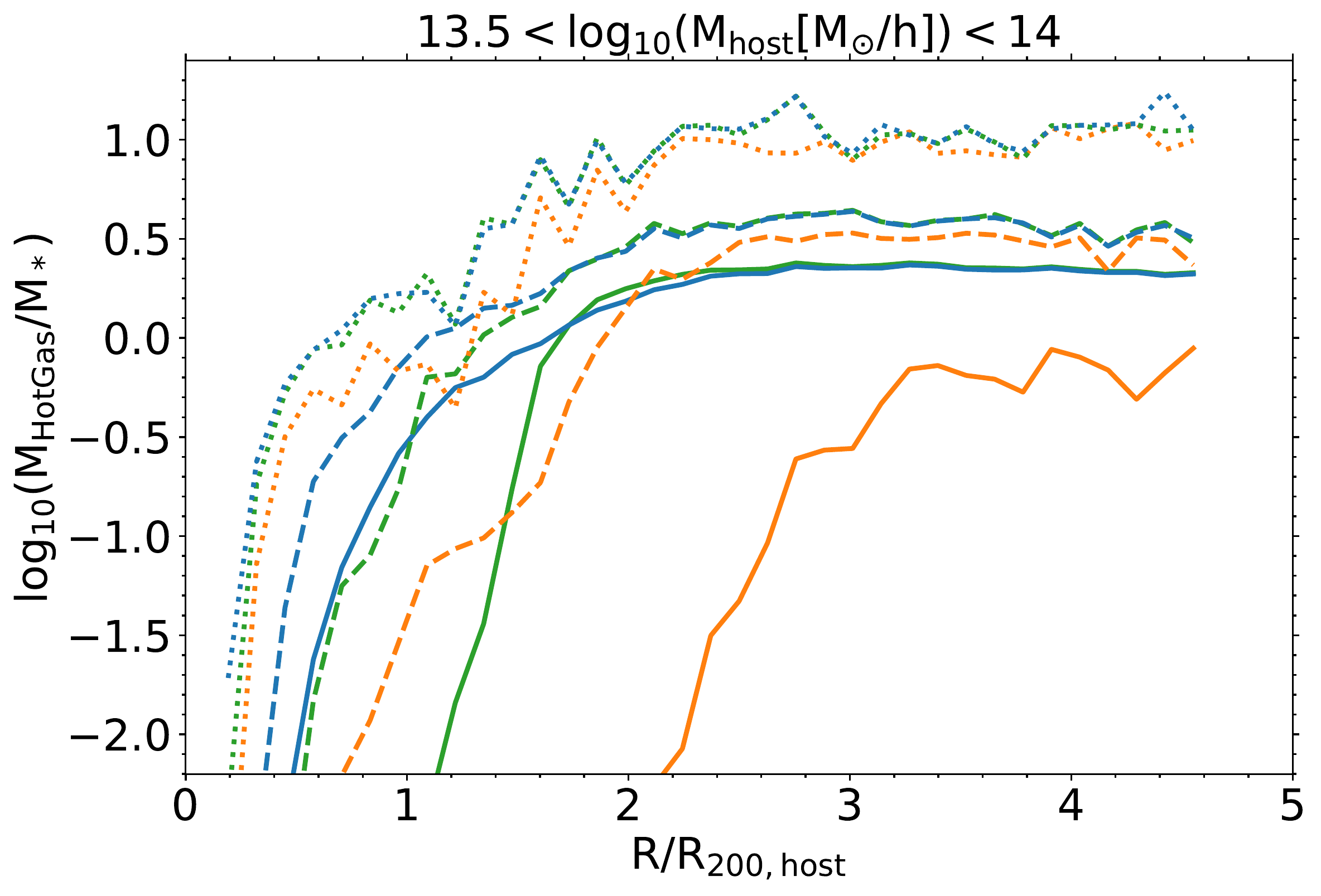}
\includegraphics[width=0.9\columnwidth]{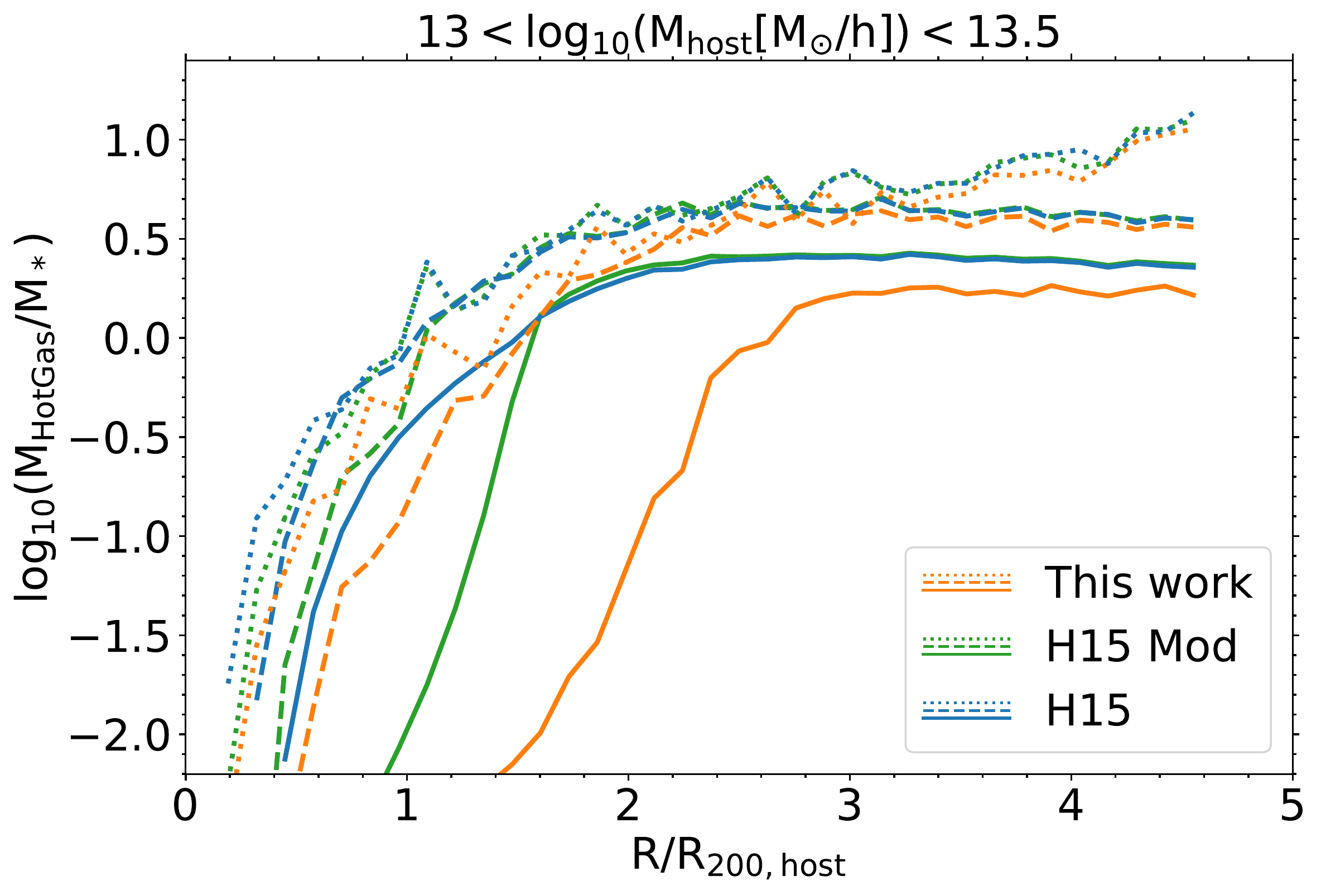}
\includegraphics[width=0.9\columnwidth]{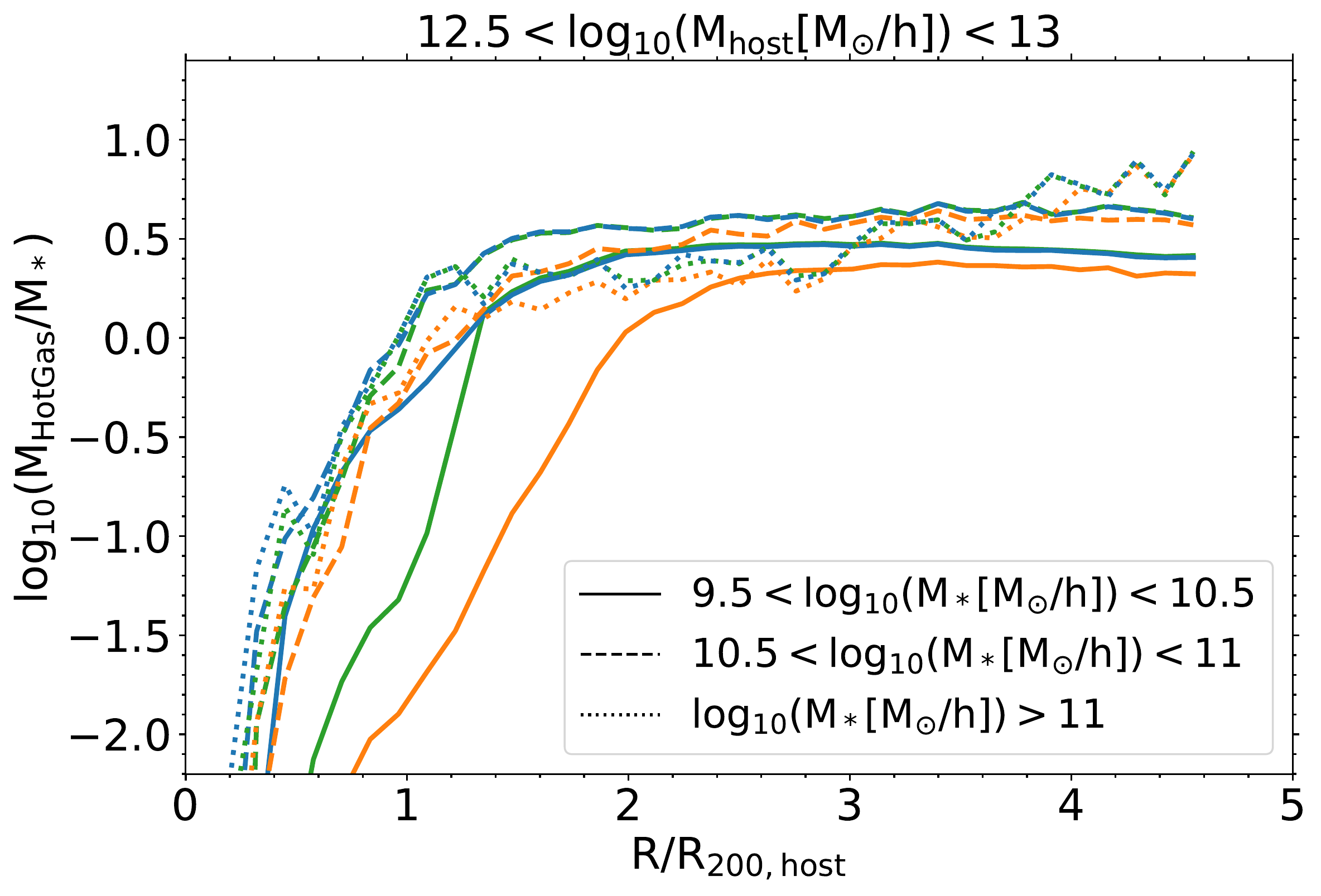}
\caption{Median value of hot gas to stellar mass ratio as a function of distance to the central host haloes. Different panels show different central host masses. In all of the panels, orange lines are the results of this work, with stripping extended to all of the galaxies in the simulation, regardless of type, position, or host mass. Blue lines are the results of H15 and green lines are the H15 modified. Different line styles correspond to different stellar mass ranges, where solid lines are for the galaxies with $\rm 9.5<log_{10}(M_*[ \rm M_{\odot}/h])<10.5$, dashed line show the ones with $\rm 10.5<log_{10}(M_*[ \rm M_{\odot}/h])<11$, and dotted lines illustrate galaxies with $\rm log_{10}(M_*[ \rm M_{\odot}/h])>11$.}
\label{fig: median_HotGasFrac_MStellar}
\end{figure*}

\subsubsection{Total stripped hot gas mass}

We assess the direct impact of the new model by looking at the total hot gas mass that a galaxy has lost due to stripping processes through the whole history of its main progenitor branch. We define the total stripped hot gas fraction as

\begin{equation}
\label{eq: f_strippedhot}
f_{\rm StrippedHot} = M_{\rm StrippedHot} \,/\, (f_{\rm b} \,M_{\rm 200,max}) \,,
\end{equation} 

\noindent where $f_{\rm b}$ is the cosmic baryon fraction and $M_{\rm 200,max}$ is the maximum value of  $M_{200}$ throughout the history of the subhalo (see \S \ref{sec: Methodology}). $M_{\rm StrippedHot}$ is total stripped hot gas mass:

\begin{equation}
M_{\rm StrippedHot} = \displaystyle\sum_{\rm z_{\rm i}}^{\rm N_{\rm snap}} m_{\rm strippedhot}(z_{\rm i}) \,,
\end{equation}

\noindent where $N_{\rm snap}$ is the number of simulation snapshots and $m_{\rm strippedhot}(z_{\rm i})$ is the sum of the mass stripped from the hot reservoir and the ejected reservoir at $z=z_{i}$. We track and record this quantity the \textsc{L-Galaxies} code, and not simply in post processing.

We analyze the total stripped hot gas fraction of galaxies, stacking systems around central host haloes out to a distance of $R=5 R_{\rm 200}$. Fig. \ref{fig: median_totStrHot_MStellar} shows the median value of $f_{\rm StrippedHot}$ as a function of distance. Each panel corresponds to a different host FOF mass range. The results are shown in three different stellar mass ranges: $9.5 < \log(M_\star [\rm M_{\odot}/h]) < 10.5$ (solid lines),  $10.5 < \log(M_\star [\rm M_{\odot}/h]) < 11$ (dashed lines), and $\log(M_\star [\rm M_{\odot}/h]) > 11$ (dotted lines). The modified H15 model (green) and the original H15 model (blue) results are shown for comparison with our new model results (orange). 

In general, our model predicts more stripping than both the H15 and H15 Mod models. The difference is small for galaxies within $R_{\rm 200}$ and becomes significant for galaxies at larger distances. For scales where $R \gtrapprox 2.5R_{\rm 200}$, more than half of the galaxies in H15 and H15 Mod have experienced no stripping. On the other hand, our model predicts stripping for galaxies to much larger scales. It can be seen that low-mass galaxies are non-trivially affected everywhere. This stripping is a combination of satellites interacting with the medium of their host haloes (so-called `pre-processing' in galaxy groups), as well as central galaxies interacting with the large-scale matter distribution of cosmic structures, e.g. sheets and filaments.

We also see that galaxies in the vicinity of more massive FOF haloes lose a larger fraction of their hot gas mass due to stripping. Comparing different panels, we conclude that more massive FOF haloes have influence on subhaloes out to larger distances -- the total stripped mass fraction decreases from the top left panel (most massive clusters) to the bottom right panel (least massive groups). As more massive haloes are found in denser environments, the increased stripping at $R/R_{200} > 2$ is likely due to ram pressure in neighboring dense haloes. Stripping is in general stronger for galaxies with lower stellar mass due to their weaker gravity, implying that more massive galaxies better retain their hot haloes.

\begin{figure*}
\includegraphics[width=0.9\columnwidth]{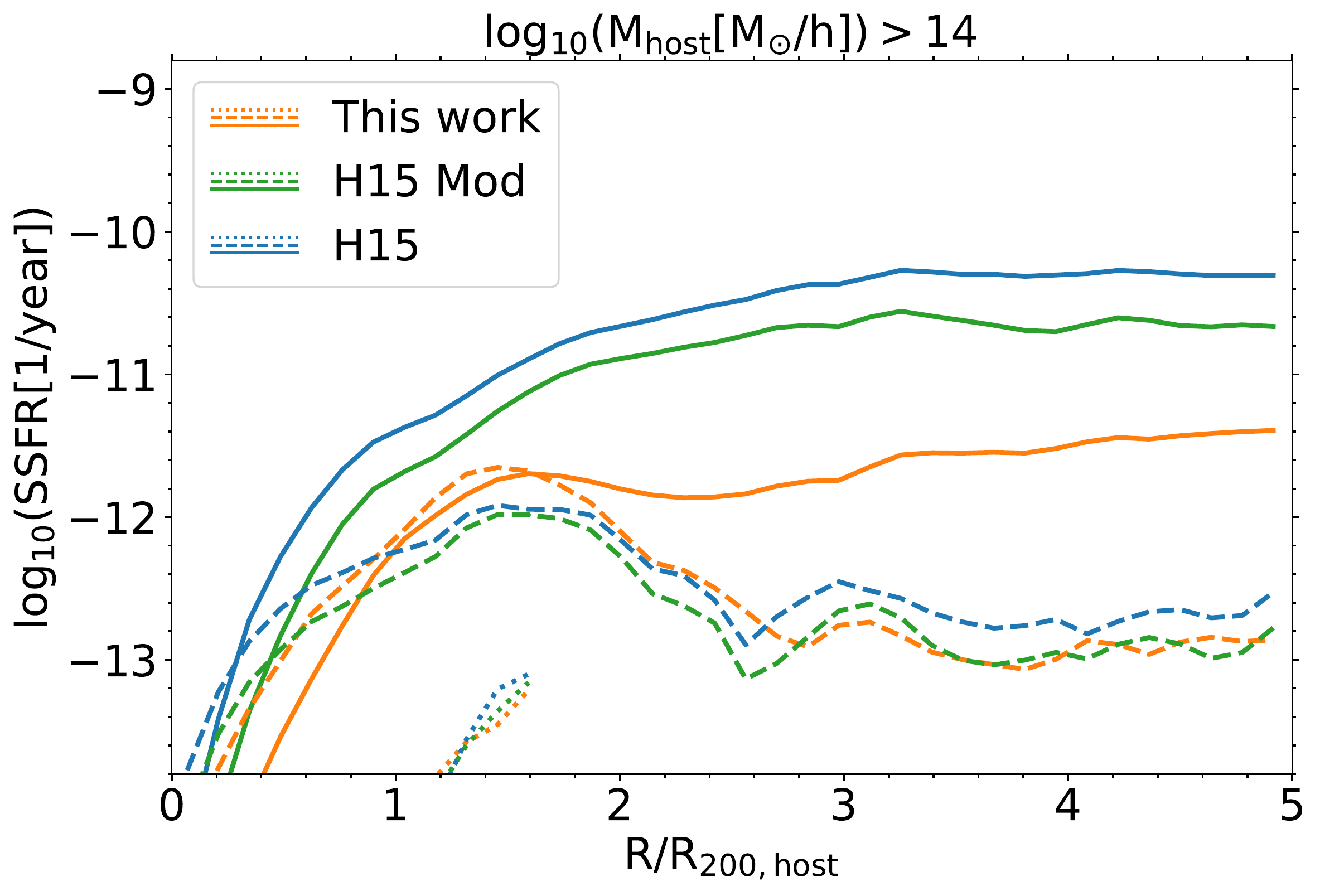}
\includegraphics[width=0.9\columnwidth]{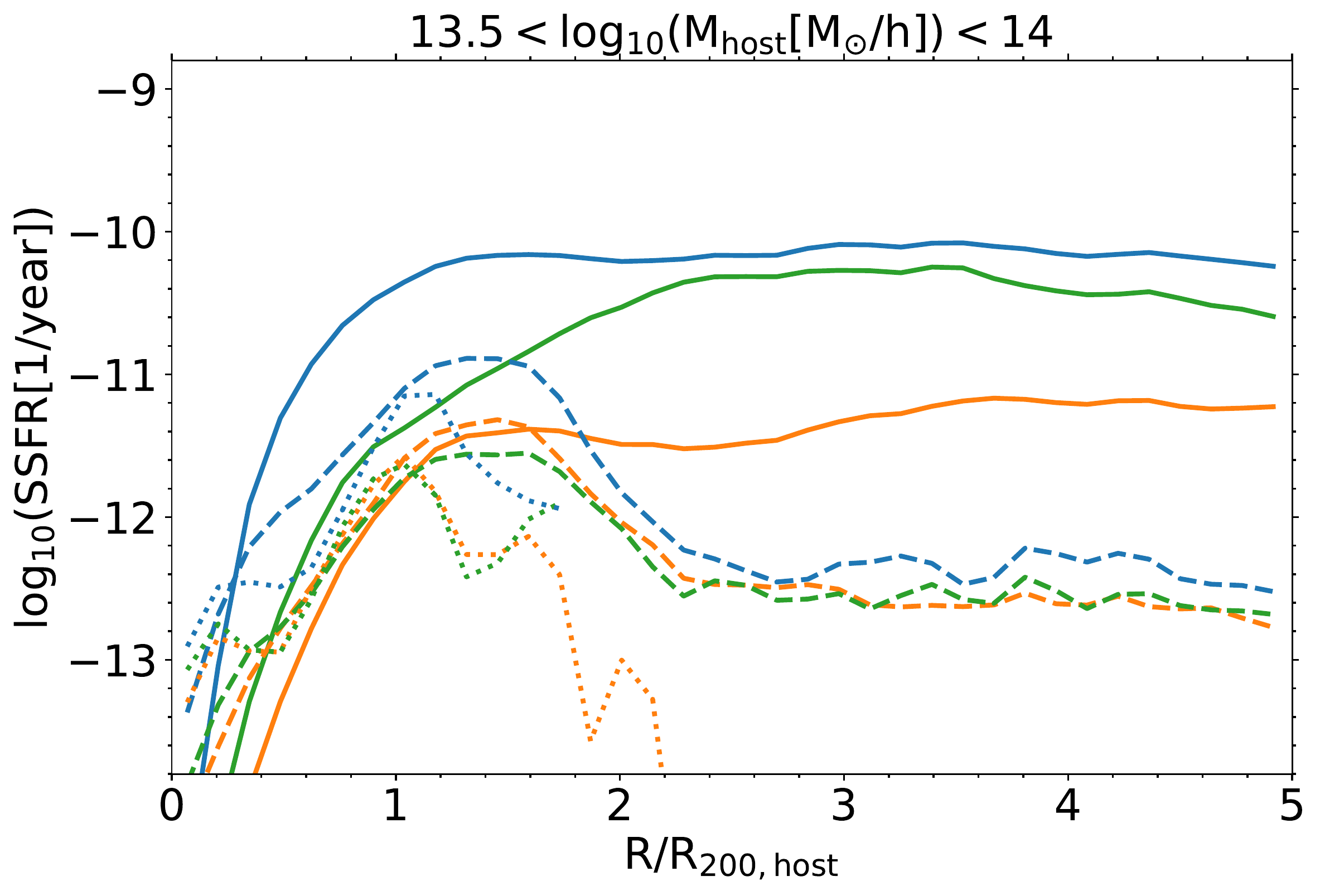}
\includegraphics[width=0.9\columnwidth]{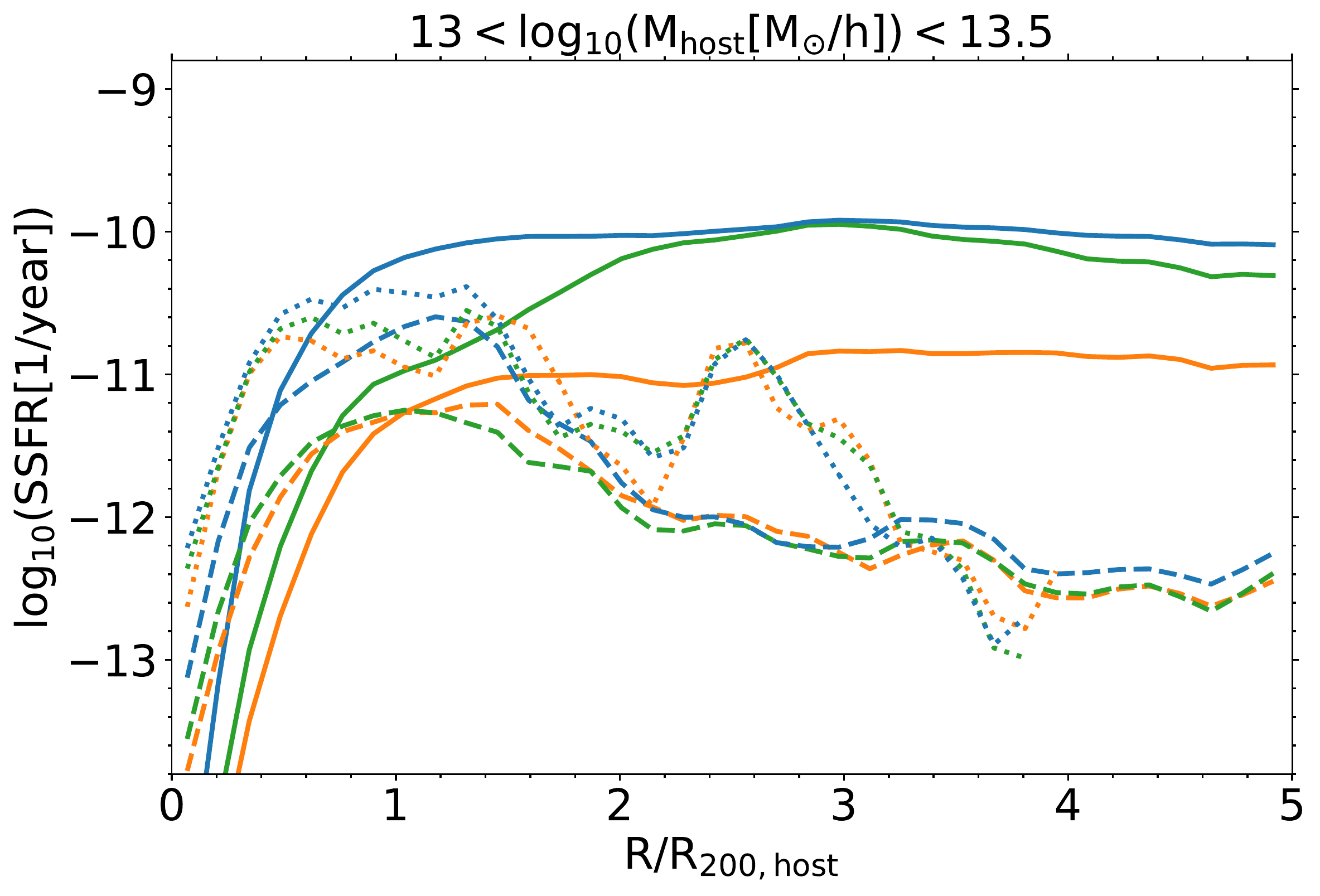}
\includegraphics[width=0.9\columnwidth]{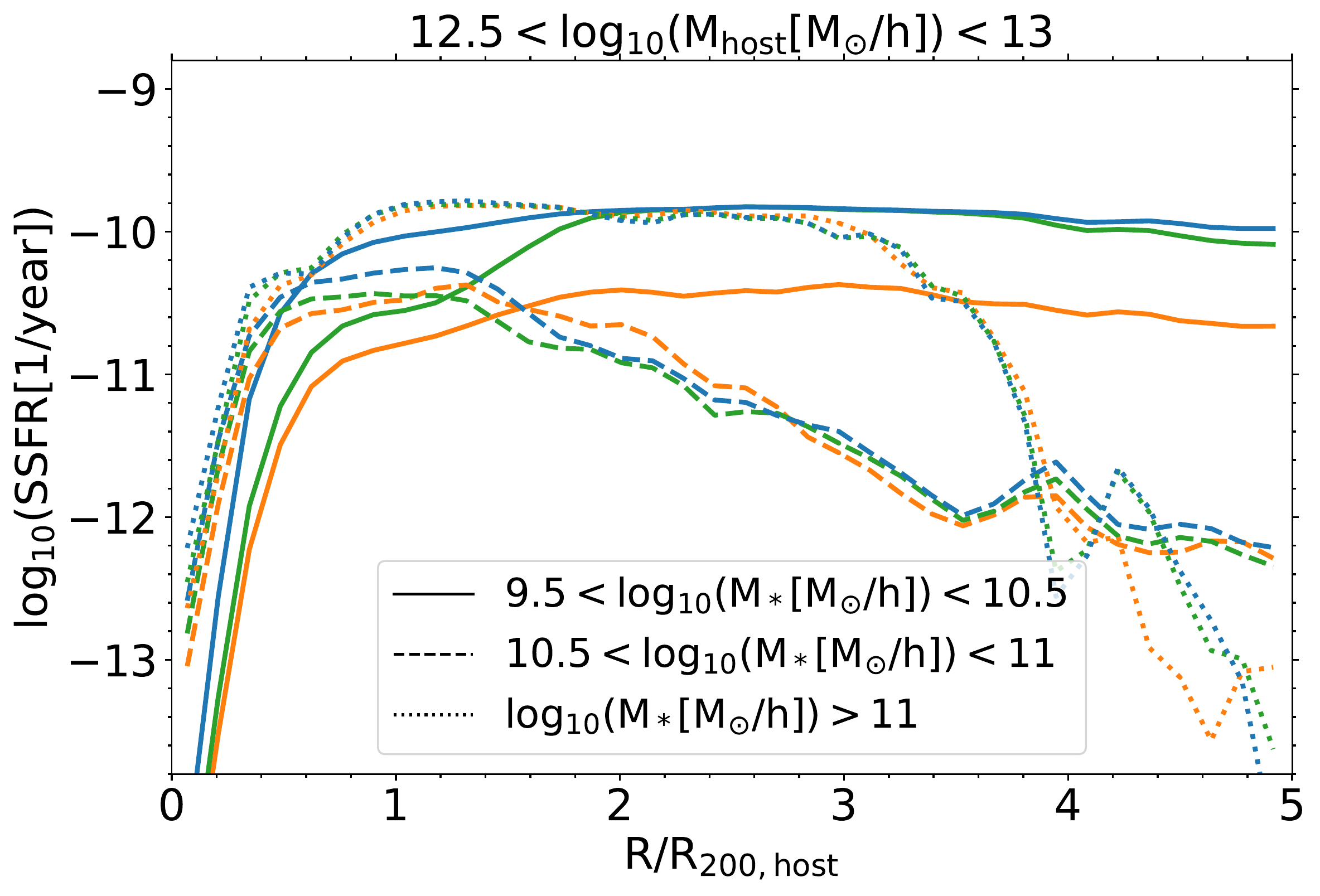}
\caption{Median value of the specific star formation rate (sSFR) of galaxies as a function of their distance from massive haloes. Different panels show different central host masses. In all of the panels, orange lines are the results of this work, with stripping applied to all of the galaxies in the simulation, regardless of type, position, or host mass. Blue lines are the results of H15 and green lines are for  the H15 modified models. Different line styles correspond to different stellar mass ranges, where solid lines are for the galaxies with $9.5 < \log(M_\star [\rm M_{\odot}/h]) < 10.5$, dashed lines are for $10.5 < \log(M_\star [\rm M_{\odot}/h]) < 11$, and dotted lines illustrate galaxies with $\log(M_\star [\rm M_{\odot}/h]) > 11$.}
\label{fig: median_SSFRGas_MStellar}
\end{figure*}

Interestingly, some stripping is seen for H15 and H15 Mod beyond $R_{200}$, even though tidal and ram-pressure stripping is not applied beyond $R_{200}$ in those models. The main cause of this phenomenon is splashback galaxies which have previously experienced stripping phases when they were inside $R_{200}$, but have since passed outside this radius. The sharp drop in the stripping fraction of H15 and H15 Mod in Fig. \ref{fig: median_totStrHot_MStellar}, is related to the splashback radius, but is not necessarily equivalent to it, because we show here the integrated stripped mass through the galaxy's entire history. We note that many of the galaxies we consider as galaxies in the vicinity of a FOF halo could have been part of other FOF haloes in the past.
\subsubsection{Hot gas fractions}
We next consider the hot gas fraction -- the ratio of hot halo gas mass to galaxy stellar mass -- in Fig. \ref{fig: median_HotGasFrac_MStellar}. As before, we stack galaxies in the vicinity of FOF haloes as a function of distance. Comparing the different models, we see that our model generically predicts a smaller hot gas to stellar mass ratio. Hot gas fractions can be suppressed by an order of magnitude or more for low mass galaxies around massive hosts. The difference between our model and H15 (or H15 Mod) is smaller for galaxies in the vicinity of low mass FOF haloes (bottom right panel) and gets larger with host FOF mass -- gas stripping is stronger near more massive haloes. Furthermore, the hot gas to stellar mass ratio changes more for galaxies with low stellar masses (solid lines) because of the weak gravitational binding energy of these systems. For galaxies with high stellar mass (dotted lines), our model approaches the previous results of the H15 and H15 Mod models. The hot gas contents of high-mass haloes ($\gtrsim 10^{13} \rm{M}_\odot$) are observationally well constrained by existing X-ray measurements (\eg{}\citealt{anderson15}), so it is again a useful sanity check that we have not significantly changed the results of \textsc{L-Galaxies} in this regime. Upcoming instruments and joint Sunyaev-Zeldovich analyses will offer tight constraints on the hot gas contents down to even lower mass scales \citep{lim18,singh18}.

In all the three models, the hot gas to stellar mass ratio first increases with distance from the centre of the FOF and then flattens. For H15 and H15 Mod, this scale is around $1.5R_{200}$, which is close to the splashback radius. On the other hand, in our model, the hot gas to stellar mass ratio flattens at a scale larger than $1.5R_{\rm 200}$ and there is no clear sign of the splashback radius. We recall that our estimation of local background density -- and so RPS -- tracks the underlying dark matter distribution and so returns continuous values as function of distance away from any host. At the same time, it captures non-spherical effects such as higher overdensities in certain directions due to the filamentary infall of satellites, a topic we will study more in the future.

\subsubsection{Specific star formation rates}

Stripping of the hot gas changes the cooling rates and the masses of cold star-forming gas in galaxies and consequently, their star formation rates. Our model generically produces lower specific star formation rates for most galaxies, as shown in Fig. \ref{fig: median_SSFRGas_MStellar}. Here we derive the median specific star formation rate (sSFR) of galaxies as a function of distance, as before. The colours and line styles are the same as in Figs. \ref{fig: median_totStrHot_MStellar} and \ref{fig: median_HotGasFrac_MStellar}. Our model (orange) reduces the sSFR for galaxies with low stellar mass (solid lines) significantly. On the other hand, the predicted values of sSFR for galaxies with $\log(M_\star [\rm M_{\odot}/h]) \gtrsim 10.5$ are not notably different from H15 or H15 Mod (dashed and dotted lines). As already seen in the hot gas fractions, low mass galaxies experience the most stripping, and the ram-pressure force at large distances is not strong enough to cause significant gas stripping in more massive subhaloes.

In contrast to hot gas contents, the star formation activity of galaxies is well studied and constrained across a wide range of mass, redshift, and environment \citep{kauffmann2004environmental,peng10,wetzel2012galaxy}. The tight star formation main sequence \citep{noeske07} and its dependence on these properties provides strong constraints on simulations \citep{wang14,donnari18}. The colour distributions and relative fractions of red versus blue galaxies, including radial member colour profiles inside groups and clusters, are similarly useful benchmarks \citep{sales15,trayford16,nelson18a}. It is clear that, in certain regimes, the sSFR changes incurred by our new RPS model updates may exceed the room allowed by current constraints. This is not surprising, as the current calibration of the \textsc{L-Galaxies} parameter set is based on the previous model \citep{henriques2015galaxy}. In the future, to make robust and realistic connections to observations, we will need to recalibrate the new model with our stripping changes.

\subsubsection{Central versus satellite galaxy properties}

Although our RPS implementation does not explicitly distinguish between central and satellite galaxies, we find as an outcome that satellites are much more influenced by RPS than centrals. At large distances from the centers of massive haloes ($R/R_{\rm 200,host} \gtrapprox 2.5$), satellites are more strongly stripped on average. This is because, by virtue of being a satellite, they typically exist in a more crowded environment. They consequently have lower hot gas to stellar mass ratios and specific star formation rates than centrals, at fixed mass.

\begin{figure*}
\centering
\includegraphics[width=0.32\textwidth]{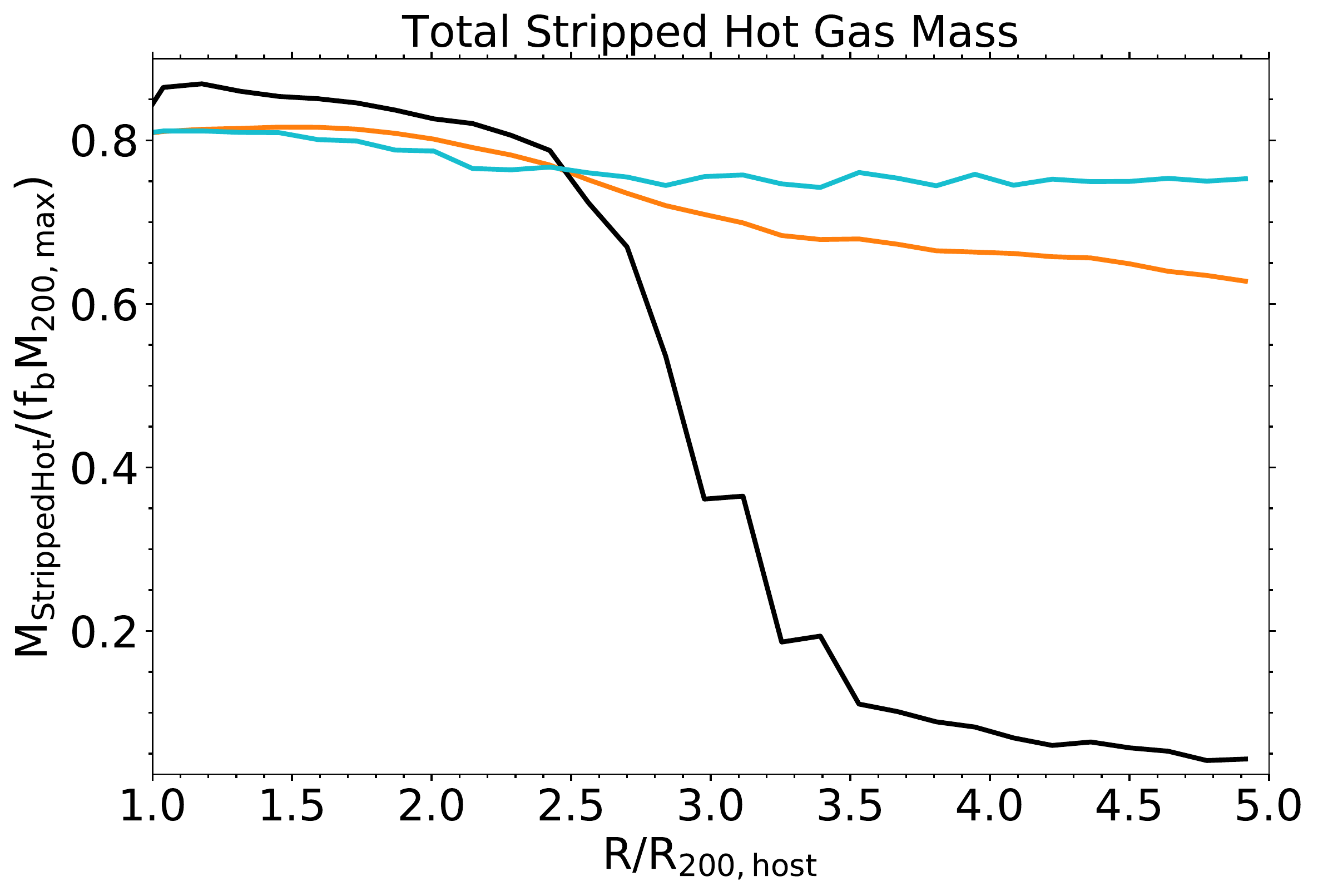}
\includegraphics[width=0.32\textwidth]{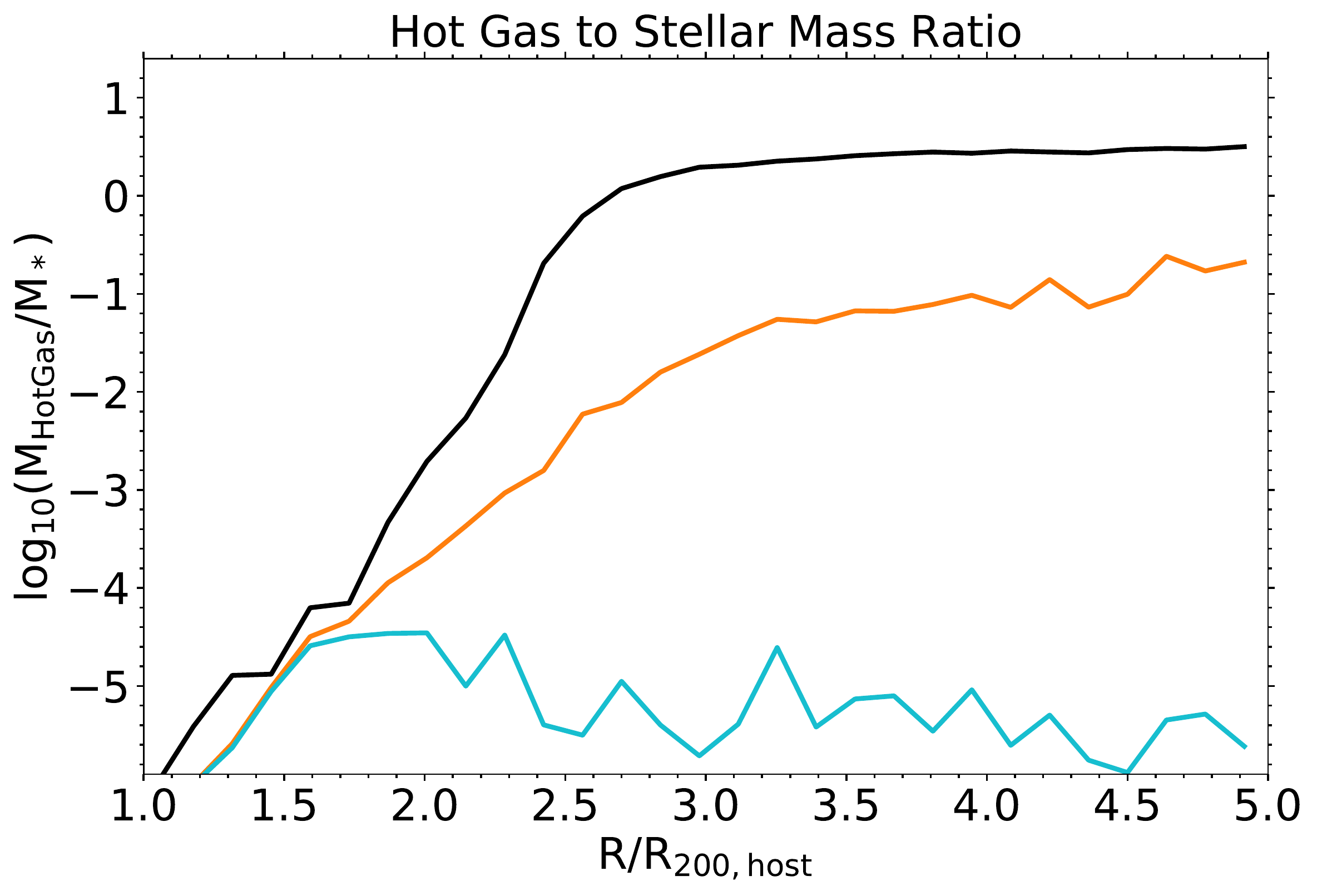}
\includegraphics[width=0.32\textwidth]{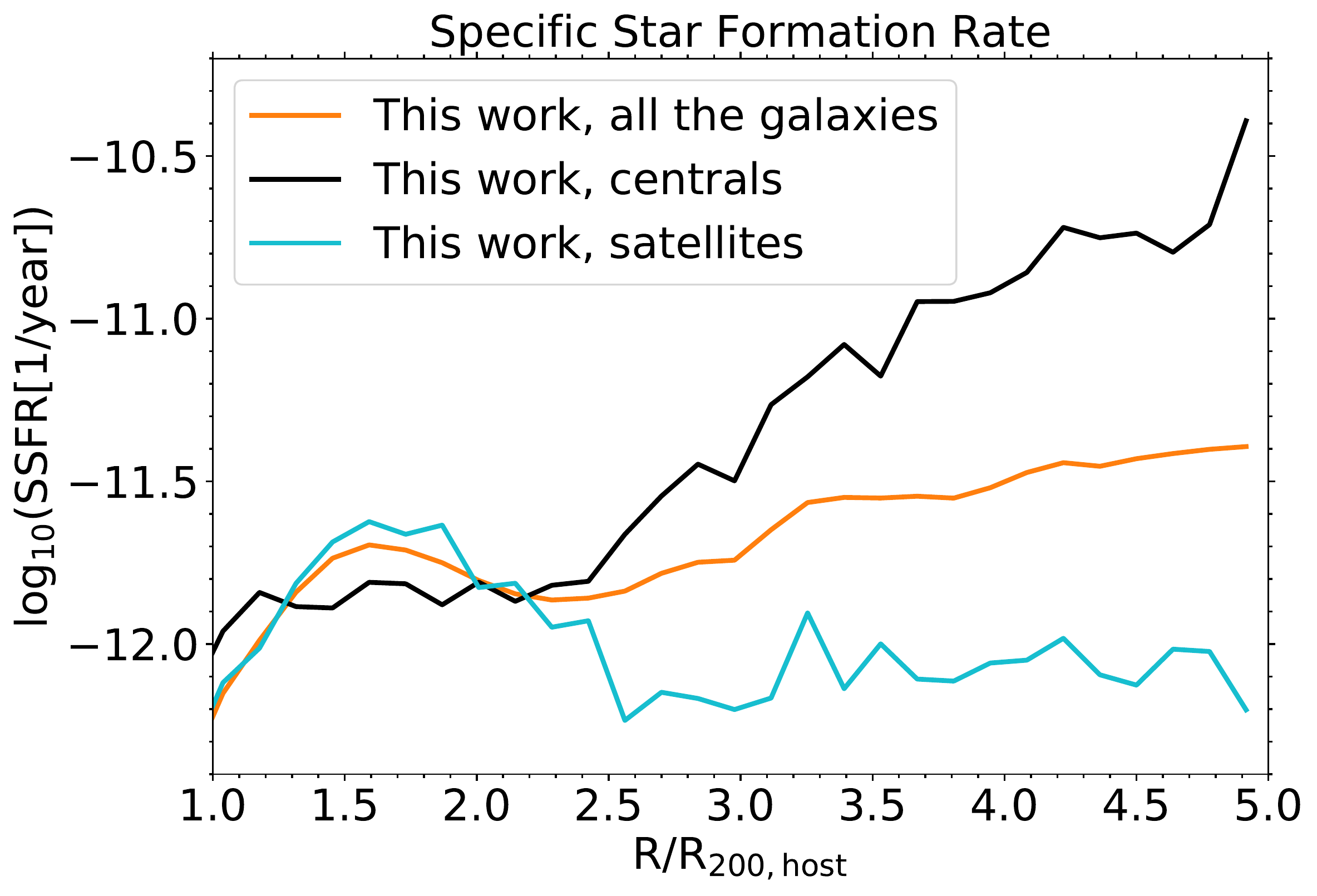}
\caption{Total stripped hot gas mass (left), hot gas to stellar mass ratio (middle), and specific star formation rate (right) of galaxies as a function of distance from a central massive halo. This figure contrasts results for central and satellite galaxies beyond $R_{200}$ in our model. Orange shows the median values for all the galaxies, while black and cyan show values separately for central and satellite galaxies. All the galaxies are located around FOF haloes with $M_{\rm 200,host} > 10^{14} \rm M_{\odot}/h$ and have stellar masses of $\rm 9.5 < log(M_\star [\rm M_{\odot}/h]) < 10.5$. }
\label{Fig: sat_cen_differences}
\end{figure*}

To clearly diagnose the role of stripping for centrals versus satellites, Fig. \ref{Fig: sat_cen_differences} shows again the total stripped hot gas fraction (left), hot gas to stellar mass ratio (middle), and sSFR (right) for satellites and centrals separately. We focus on the most massive host halo bin, $M_{\rm 200,host} > 10^{14} \rm{M}_\odot/\rm{h}$ (top left panels of Figs. \ref{fig: median_totStrHot_MStellar} through \ref{fig: median_SSFRGas_MStellar}) and concentrate on the outer regions of haloes, beyond $R_{\rm 200}$. We decompose the median for all galaxies (orange) into centrals (black) and satellites (cyan) separately.

As can be seen in the left panel, at large radii, satellite galaxies have lost a significant fraction of their hot gas due to stripping, while central galaxies have lost only a small fraction. Specifically, satellites have a stripped fraction above 75\% regardless of distance, while this fraction drops rapidly to zero for centrals beyond $\sim 3 R_{\rm 200}$. This constancy for satellite systems hightlights the important role of environmental effects associated to host haloes other than the central one around which we stack. 

The middle panel of Fig. \ref{Fig: sat_cen_differences} makes it clear that, at large radii and for galaxies with $10^{9.5} < M_\star / (\rm M_\odot/h) < 10^{10.5}$, the hot gas to stellar mass ratio of central galaxies is much higher than in satellites, exceeding unity in general. In our fiducial model, satellites at these stellar masses have lost the vast majority of their hot haloes, and have negligible hot gas fractions. Once centrals start to experience similarly harsh environments ($R/R_{\rm 200} \lesssim 2$) they become equally depleted of their hot gaseous reservoirs. Galaxies, regardless of type or history, cannot retain hot halo gas once they approach such massive hosts.

The right panel of Fig. \ref{Fig: sat_cen_differences} shows that central galaxies have higher specific star formation rates at large radii compared to satellites. This differential effect can exceed 1 dex in sSFR, although an equal level of quiescence is reached by the time galaxies approach $R/R_{\rm 200} \simeq 2.5$ (i.e. $\sim 3$ Mpc) from the FOF halo, demonstrating the importance of large-scale environmental effects far beyond the virial radii of massive haloes.


\section{Summary}
\label{sec: conclusions}

In this work we have introduced a new technique to capture environmental effects, in particular, ram-pressure stripping (RPS), within semi-analytical models of galaxy formation. To do so we have introduced the concept of the local background environment (LBE). The LBE is extracted directly from the particle-level data of the high resolution N-body cosmological simulation of structure formation on which \textsc{L-Galaxies} is run, using a local, adaptive, spherical shell. We design a Gaussian mixture estimator to separate true background particles from subhalo contaminants in velocity space, overcoming classical difficulties of substructure finders in dense environments.

We first measure the statistical properties of the local background environment of all subhaloes in the Millennium Simulation, analyzing its properties at $z=0$. We find that:

\begin{itemize}

\item Neither the LBE density ($\rho_{\rm LBE}$) nor the galaxy's velocity relative to its LBE ($\vec{v}_{\rm gal,LBE}$) show a strong dependence on subhalo mass. The LBE density of satellite galaxies does not vary significantly with host halo mass, but the velocity of satellite galaxies relative to their LBE strongly increases with host mass. Central galaxies move faster relative to their LBE with increasing mass. The velocity of a galaxy relative to its environment, $\vec{v}_{\rm gal,LBE}$, declines slowly with increasing distance away from massive hosts.
 
\item Both $\rho_{\rm LBE}$ and $v_{\rm gal,LBE}$  vary continuously and weakly across the virial radius of the host halo. Neither exhibit discontinuous behavior out to $5R_{\rm 200}$, implying that ram-pressure stripping  extends (and can be non-negligible) beyond the virial radius.

\item At large distances from massive haloes, the LBE of galaxies moves on average in the same direction as the galaxy itself, although typically somewhat slower, a signature of coherent infall. The angle between $\vec{v}_{\rm gal}$ and $\vec{v}_{\rm LBE}$ becomes more uniform towards halo centers as orbits isotropize. The LBE of satellite galaxies is generically \textit{not} at rest with respect to the host halo, in contrast to common assumptions.

\end{itemize}

Using our LBE methodology we then devise a new treatment of ram-pressure stripping (RPS) of hot halo gas within the Munich semi-analytical model \textsc{L-Galaxies}. Applied to the Millennium simulation, our principal results are:

\begin{itemize}
\item Compared to the publicly released version of \textsc{L-Galaxies} \citep{henriques2015galaxy} our model results in significantly more hot gas stripping. While the stellar mass function remains unchanged, the hot gas fractions and specific star formation rates are strongly suppressed, depending on satellite mass, host mass, and distance.

\item Galaxies with the lowest stellar masses are the most affected by stripping due to their weaker self gravity. Satellites in massive clusters ($M_{\rm host} > 10^{14} \rm{M}_\odot$) with $M_\star < 10^{10.5} M_\odot$ can lose the majority of their hot gas mass relative to the fiducial model, while for $M_\star > 10^{11} M_\odot$ the impact is subtle.

\item Our LBE estimates imply that galaxies near and inside group-mass haloes ($< 10^{14} \rm{M}_\odot$) also undergo sufficient RPS to impact their hot gas contents and so star formation rates, despite lower background densities and relative velocities.

\item Satellites at large distances ($> 2 R_{\rm 200}$) experience much stronger RPS effects than centrals at the same mass, indicative of environmental pre-processing.

\end{itemize}

In this work we have not yet compared the new model outcomes with observational data. This will require a recalibration of the free parameters in \textsc{L-Galaxies} \citep[as in][]{henriques2015galaxy}, which we will undertake in the future. Several assumptions of our treatment for stripping, such as the mapping between the dark matter and gas matter density fields, will also benefit from comparisons with hydrodynamical simulations \citep{nelson19a}. Overall, our results demonstrate the importance of a ram-pressure stripping model which incorporates local (and \textit{continuous}) estimates of background properties without artificial boundaries in space or halo mass.

\section*{Acknowledgements}

MA thanks Volker Springel, Stefan Hilbert, and Wolfgang Enzi for useful discussions and assistance.

\vspace{-1em}
\bibliographystyle{mnras}
\bibliography{refbibtex}

\bsp
\label{lastpage}

\end{document}